\documentclass[journal]{IEEEtran}

\ifCLASSINFOpdf
\usepackage[pdftex]{graphicx}
\DeclareGraphicsExtensions{.pdf,.jpeg,.png}
\else
\usepackage[dvips]{graphicx}
\fi
\usepackage{subfigure}
\usepackage{epstopdf}
\usepackage[cmex10]{amsmath}
\usepackage{amssymb}
\usepackage{wasysym}
\usepackage{relsize}
\usepackage{algorithm}
\usepackage{algorithmic}
\usepackage{array}
\usepackage{cite}
\usepackage{color}
\usepackage{url}
\usepackage{bm}
\usepackage{enumerate}
\usepackage{epsfig,latexsym}
\usepackage{cite}

\begin{document}
\title{DBRAA: Sub-6 GHz and Millimeter Wave Dual-Band Reconfigurable Antenna Array for ISAC}

\author{Kangjian~Chen,~\IEEEmembership{Student~Member,~IEEE}, Chenhao~Qi,~\IEEEmembership{Senior~Member,~IEEE},  and Octavia A. Dobre,~\IEEEmembership{Fellow,~IEEE}
	
	\thanks{ This work was supported in part by	the National Natural Science Foundation of China. Part of this work is to be submitted to the IEEE Global Communications Conference, Taipei, Taiwan, Dec. 2025~\cite{GlobeCom25CKJ}. (\textit{Corresponding author: Chenhao~Qi})}
	\thanks{Kangjian~Chen and Chenhao~Qi are with the School of Information Science and Engineering, Southeast University, Nanjing 210096, China (e-mail: \{kjchen,qch\}@seu.edu.cn).}
	\thanks{Octavia A. Dobre is with the Faculty of Engineering and Applied Science, Memorial University, St. John’s, NL A1C 5S7, Canada (e-mail: odobre@mun.ca).}
}

{}

\maketitle

\begin{abstract}
This paper proposes a dual-band reconfigurable antenna array (DBRAA),  enabling wireless capabilities in both sub-6 GHz (sub-6G) and millimeter wave (mmWave) bands using a single array.  For the sub-6G band, we  propose a reconfigurable antenna selection structure, where each  sub-6G antenna is formed by multiplexing several mmWave antennas, with its position dynamically adjusted using PIN diodes. For the mmWave band, we develop a reconfigurable hybrid beamforming structure that connects radio frequency chains to the antennas via phase shifters and a reconfigurable switch network. We then investigate integrated sensing and communications (ISAC) in sub-6G and mmWave bands using the proposed DBRAA and formulate a dual-band ISAC beamforming design problem. This problem aims at maximizing the mmWave communication sum-rate subject to the constraints of sub-6G communication quality of service and sensing beamforming gain requirements. The dual-band ISAC beamforming design is decoupled into sub-6G beamforming design and mmWave beamforming design. For the sub-6G beamforming design, we develop a fast search-based joint beamforming and antenna selection algorithm. For the mmWave beamforming design, we develop an alternating direction method of multipliers-based reconfigurable hybrid beamforming algorithm. Simulation results demonstrate the effectiveness of the proposed methods.

\end{abstract}



\section{Introduction}
The sixth-generation  (6G) wireless communications have attracted  widespread interest from both academia and industry. With its potential to facilitate unprecedented data rates, low latency, and massive connectivity, 6G is expected to promote the development of numerous applications, including telemedicine, autonomous vehicles, smart homes, and industrial intelligence~\cite{yang2019,OBETrans}. One of the most promising technologies for fulfilling the visions of  6G is radar sensing~\cite{IoT2024LSH},  which enables 6G networks to achieve high-accuracy localization, object detection, recognition, and imaging. By seamlessly integrating sensing into communication networks, 6G can meet the stringent requirements of various  compelling applications. For instance, 6G-enabled sensing can facilitate real-time monitoring of the environment, enabling efficient management of urban infrastructure and public services~\cite{Tcom24WZQ}. To address the critical requirement for sensing in 6G, integrated sensing and communications (ISAC) has been developed~\cite{Tcom20Liufan}. Initially, sensing and communications operated as separate entities~\cite{JSTSP18Zhengle},  resulting in high hardware costs and inefficient resource utilization. Considering the substantial similarities  between sensing and communications in terms of antennas, radio frequency (RF) devices, frequency bands, and processing algorithms, integrating the two systems into a unified one has been a focus~\cite{TWC23CKJ1,IEEENetwork24_DFW,TWC23DFW}. This integration enables wireless networks to perform sensing and communication tasks with substantially reduced hardware and software resources, thereby achieving integration gains. Furthermore, integrating the two systems into a single entity facilitates seamless cooperation and enhances the performance of both functions~\cite{CL24CKJ}, leading to coordination gains. For instance, in vehicle-to-everything networks, the communication systems can leverage sensing results and kinematic models for more efficient beam tracking~\cite{TWC24WY}. Additionally, the performance of sensing systems  can be enhanced through cooperation within ubiquitous communication networks. The significant benefits of integration and coordination gains have spurred extensive research and development in ISAC.
 

The sub-6 GHz (sub-6G) and millimeter wave (mmWave) frequency bands  are essential for fulfilling the demands of future wireless networks~\cite{JSAC17SM}. The mmWave band, known for its high data rates based on abundant spectrum resources, is particularly suited for dense urban environments and applications requiring large data throughput. However, it suffers from high path loss, restricting its effective coverage. Conversely, the sub-6G band offers better penetration and wider coverage, making it more suitable for wide area connectivity. However, its data rates may be lower than the mmWave band due to limited spectrum resources. Additionally, these bands also exhibit different sensing capabilities. The mmWave band, with its shorter wavelength and larger numbers of antennas, can achieve higher-precision sensing than the sub-6G band~\cite{CL22Chenhao}. On the other hand, the sub-6G band, with lower path loss, can provide more reliable sensing than the mmWave band in complicated environments. The complementary characteristics of mmWave and sub-6G bands make them an ideal combination for future 6G networks~\cite{TVT20CM}. By harnessing the strengths of both  bands, 6G networks can leverage the high-rate communication and high-accuracy sensing capabilities of the mmWave band while ensuring extensive coverage and reliability with the sub-6G band. Therefore, the development of a dual-band ISAC framework is significant for unlocking the full potential of 6G systems.



Another perspective to improve the performance of wireless communications is the evolution of antenna techniques. One notable technique is the dual-band antenna array. The significant frequency difference between sub-6G and mmWave bands results in vastly different antenna sizes, as antenna sizes are directly related to the carrier wavelength~\cite{Balanis2015}. As a result, systems handling both sub-6G and mmWave bands typically require two separate antenna arrays, leading to significant physical space occupation and hardware costs. To address this issue, dual-band antennas that can operate in both sub-6G and mmWave bands have been developed~\cite{TAP23_SL,AWPL24TXY}. These dual-band antennas combine multiple mmWave antennas into a single sub-6G antenna, allowing for shared apertures between the two frequency bands and resulting in substantial space savings.  However, this method offers limited design flexibility, as the antenna positions cannot dynamically adapt to the varying channel state information (CSI). One method to improve the design flexibility of the antenna arrays is using antenna selection techniques, which can dynamically select a subset of antennas from a large array based on current CSI and system requirements~\cite{liu2023joint}. By selectively activating antennas according to the CSI, antenna selection enhances signal quality, reduces interference, and improves the overall system efficiency. Nevertheless, this method requires a large number of candidate antennas, leading to  high hardware costs. Additionally, the spacing between adjacent antennas is typically larger than half a wavelength to avoid coupling effects, which limits the design flexibility for antenna position optimization. To further enhance the design flexibility, two techniques, movable antennas~\cite{TWC24MWY} and fluid antennas~\cite{TWC21KKW}, have been developed. Movable antennas adjust their positions dynamically using precise mechanical systems to optimize signal reception and transmission. Fluid antennas, on the other hand, use materials like liquid metals to reconfigure their shape and position dynamically, allowing for efficient space utilization and performance optimization. However, both techniques depend on high-precision control mechanisms, which can be complicated and costly to implement. In addition to the aforementioned techniques, reconfigurable antennas present another promising approach~\cite{WC24_YKK,APM13_HR,IEEE_15_P}. These antennas can dynamically alter their electrical characteristics, such as radiation pattern and polarization, to meet diverse requirements. By utilizing variable RF components like PIN diodes, switches, and tuners, reconfigurable antennas can achieve multifunctional operations without changing their physical structure. This adaptability allows reconfigurable antennas to maintain robust performance across various environments and applications. 


In this paper, we investigate the ISAC systems that operate in sub-6G and mmWave bands. Inspired by the existing antenna techniques,  we first propose a sub-6G and mmWave dual-band reconfigurable antenna array (DBRAA) to enable  wireless capabilities and performance improvement in both bands. Then, sub-6G and mmWave beamforming designs based on the DBRAA are developed. Our contributions are  summarized as follows.

\begin{itemize}
\item We propose a sub-6G and mmWave DBRAA, enabling wireless capabilities in both sub-6G and  mmWave bands using a single array. For the mmWave band, the DBRAA utilizes an mmWave array for signal transmission. To reduce the hardware costs, we propose a reconfigurable hybrid beamforming (RHB) structure, where the RF chains are connected to the antennas through phase shifters and a reconfigurable switch network. For the sub-6G band,  we propose a reconfigurable antenna selection (RAS) structure,  where each antenna is formed by multiplexing multiple mmWave antennas and the positions of sub-6G antennas can be dynamically adjusted by controlling the PIN diodes between mmWave antennas.

\item We investigate the sub-6G and mmWave dual-band ISAC using the proposed DBRAA. Since the sub-6G band supports reliable sensing and communications while the mmWave band supports high-rate communications and high-resolution sensing, we combine their merits and formulate a sub-6G and mmWave dual-band ISAC beamforming design problem, which aims at maximizing the mmWave communication sum-rate  subject to the constraints of the sub-6G communications quality of service (QoS)  and the sensing beamforming gain requirements. Then, the formulated dual-band ISAC beamforming design is further decoupled into the sub-6G beamforming design and the mmWave beamforming design.

\item We develop a fast search-based joint beamforming and antenna selection (FS-JBAS) algorithm for the sub-6G beamforming design. We consider a subproblem of the formulated sub-6G beamforming design, and  propose an alternating beamforming and antenna selection (ABAS) algorithm to solve the subproblem for the initialization of the FS-JBAS algorithm. Then, we iteratively  select each of the sub-6G antennas via a fast search method.

\item We develop an alternating direction method of multipliers-based RHB (ADMM-RHB) algorithm for the mmWave beamforming design. We aim at maximizing the mmWave communication sum-rate subject to the constraints of sensing beamforming gain requirements. Since the objective of sum-rate maximization  is nonconvex, the weighted mean squared error (WMMSE) is adopted to convert the objective  into an equivalent and tractable form. To deal with the constraints of nonconvex sensing beamforming gain requirements, a successive convex approximation method is developed. Then, the ADMM-RHB algorithm alternately  optimizes the variables to design the mmWave beamforming.

\end{itemize}

The rest of this paper is organized as follows. The model of the sub-6G and mmWave  dual-band  ISAC system is introduced in Section~\ref{SystemModel}. The  dual-band  ISAC beamforming design problem is formulated in Section~\ref{ProblemFormulation}. The sub-6G and mmWave beamforming designs are developed in Section~\ref{SSD} and Section~\ref{MSD}, respectively.  The proposed methods are evaluated in Section~\ref{SR}, and the paper is concluded in Section~\ref{CC}.

\begin{figure}[t]
	\centering
	\includegraphics[width=69mm]{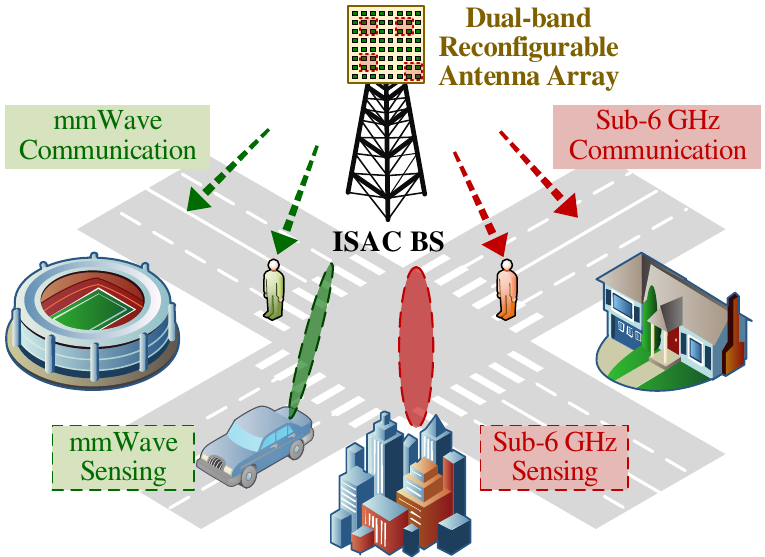}
	\caption{Illustration of the sub-6G and mmWave dual-band ISAC systems.}
	\label{fig_image1}
\end{figure}

\textit{\textbf{Notations:}} Lowercase and uppercase bold symbols denote vectors and matrices, respectively. $[\boldsymbol{A}]_{m,:}$,  $[\boldsymbol{A}]_{:,n}$, and $[\boldsymbol{A}]_{m,n}$ denote the $m$th row, the $n$th column, and the entry on the $m$th row and the $n$th column of a matrix $\boldsymbol{A}$.  $[\boldsymbol{a}]_m$ denotes the $m$th entry of a vector $\boldsymbol{a}$. $\|\boldsymbol{a}\|_2$ denotes the $\ell_2$-norm of the vector $\boldsymbol{a}$ and $\|\boldsymbol{A}\|_{\rm F}$ denotes the Frobenius norm  of the matrix $\boldsymbol{A}$. $\mathrm{vec}\{\cdot\}$ denotes the vectorization of a matrix. $(\cdot)^{\rm T}$, $(\cdot)^{\rm H}$, and $\text{diag}\{\boldsymbol{a}\}$ denote transpose, Hermitian transpose, and a matrix with vector $\boldsymbol{a}$ as its diagonal entries, respectively. $\mathbb{Z}$,  $\mathbb{C}$, $\mathcal{R}\{\cdot\}$, and $\mathcal{CN}(\cdot)$ denote  integers, complex numbers, the real part of a complex number, and complex Gaussian distribution, respectively.


\section{System Model}\label{SystemModel}
As shown in Fig.~\ref{fig_image1}, we consider an ISAC system, where one ISAC base station (BS) communicates with users and senses targets in both the sub-6G and mmWave bands. For the sub-6G band, the ISAC BS provides high-reliability communications for $K_{\rm s}$ sub-6G users while detecting $T_{\rm s}$ potential targets. For the mmWave band, the ISAC BS provides high-rate communications for $K_{\rm m}$ mmWave users and high-accuracy sensing for $T_{\rm m}$  targets. The carrier frequencies of the sub-6G   and mmWave bands are denoted as $f_{\rm s}$ and $f_{\rm m}$, respectively. Correspondingly, their carrier wavelengths are denoted as $\lambda_{\rm s}$ and $\lambda_{\rm m}$, respectively. In this work, we focus on the processing at the BS. Therefore, for simplicity, we assume that each user is equipped with a single antenna.

\begin{figure}[!t]
	\centering
	\includegraphics[width=70mm]{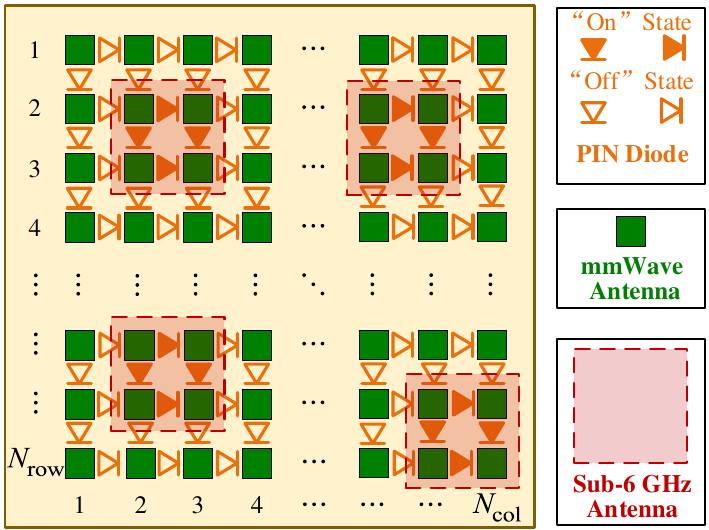}
	\caption{Illustration of the dual-band reconfigurable antenna array.}
	\label{fig_DBRAA}
\end{figure}

\subsection{Dual-Band Reconfigurable Antenna Array}\label{Dual_bandReconAA}
To enable the ISAC in the sub-6G and mmWave bands, we propose a DBRAA, which is composed of co-aperture sub-6G and mmWave antenna arrays, as shown in Fig.~\ref{fig_DBRAA}. 

\subsubsection{Overview} The DBRAA includes $N_{\rm m}$ mmWave antennas and $N_{\rm s}$ sub-6G antennas. The mmWave antennas are arranged in a half-wavelength-interval uniform planar array, including $N_{\rm row}$ rows and $N_{\rm col}$ columns. As a result,  we have $N_{\rm m} = N_{\rm row}\times N_{\rm col}$. Based on the dual-band antenna arrays designed in~\cite{TAP23_SL,AWPL24TXY}, we combine mmWave antennas to form sub-6G antennas, which  enables the co-aperture sub-6G and mmWave antenna arrays. Following \cite{AWPL24TXY}, we assume that  a single sub-6G antenna is formed by combining $2\times2$ mmWave antennas, as illustrated in Fig.~\ref{fig_DBRAA}. Note that this work can  e extended to other antenna configurations by modifying the system and signal models.  Inspired by reconfigurable pixel antennas~\cite{zhang2024pixel,TAP22_JLW,TAP17_LP}, which dynamically adjust connections between elementary pixel antennas using PIN diodes to achieve diverse electrical characteristics, we incorporate PIN diodes into the existing dual-band antenna arrays~\cite{TAP23_SL,AWPL24TXY} to dynamically adjust the connections between mmWave antennas, as illustrated in Fig.~\ref{fig_DBRAA}. By controlling the states of the PIN diodes, a sub-6G antenna can be formed by choosing  $2\times2$ mmWave antennas from the mmWave antenna array, which enables us to dynamically adjust the positions of sub-6G antennas. In addition, the inherent orthogonality between mmWave and sub-6G bands enables concurrent operation of both antenna types within the shared aperture.

\begin{figure}[!t]
	\centering
	\includegraphics[width=80mm]{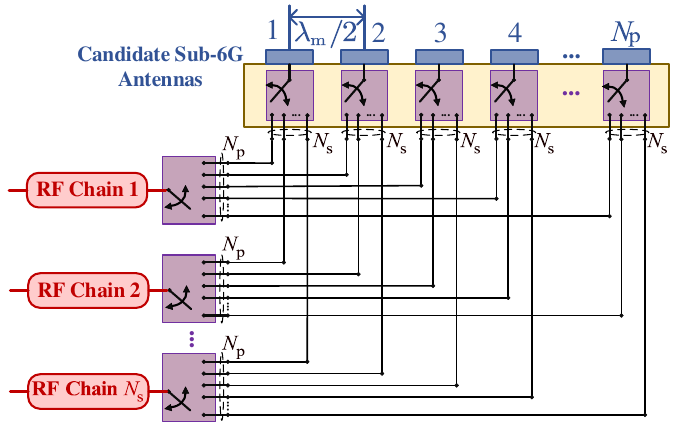}
	\caption{Illustration of the reconfigurable antenna selection structure.}
	\label{fig_AntennaSelection}
\end{figure}

\textbf{Remark 1:} The feasibility of fabricating the  DBRAA is supported by three key facts. First, in~\cite{Rao2024}, a prototype of a shared-aperture dual-band sub-6G  and mmWave reconfigurable intelligent surface has been fabricated, which indicates that integrating mmWave elements to form sub-6G antennas with the DBRAA configuration can enable independent operation in both bands. Second, the application of PIN diodes to control the interconnections between RF elements is a well-established technique that has been validated in numerous studies~\cite{TAP14HSV,Saifullah2022,TAP24ZLH,Hong2024}. Therefore, integrating the PIN diodes into the dual-band sub-6G  and mmWave reconfigurable intelligent surface in~\cite{Rao2024} to control the connections between mmWave elements is straightforward. Third, characteristic mode analysis reveals that the electromagnetic properties of RF devices are predominantly determined by their intrinsic structure rather than by the excitation mechanisms~\cite{APM15VM,TAP17LF,APM22AJJ}. Consequently, design strategies validated on reconfigurable intelligent surfaces can be systematically extended to antenna systems, as both share common underlying electromagnetic modes~\cite{TAP20WJF,APM22MD}. These facts indicate that the DBRAA can be fabricated with the dedicated efforts of antenna researchers.

\textbf{Remark 2:} In the DBRAA, some mmWave antennas are reused to form sub-6G antennas, while others remain unused. The presence of these unused metal antenna elements may degrade the performance of the sub-6G band. To address this issue, additional RF design techniques, such as the suspended electromagnetic band-gap (EBG) structure developed in \cite{Rao2024}, can be integrated into the DBRAA. As demonstrated in \cite{Rao2024}, with the EBG structure, the surface waves are confined with each element, and the mutual interference between adjacent elements is effectively mitigated.

\textbf{Remark 3:} In the DBRAA, the use of diodes to connect the mmWave antennas for forming sub-6G antennas may affect the performance of the  mmWave array. To avoid this effect,  additional RF components, such as the planar spiral inductor structure in~\cite{Rao2024}, can be designed and cascaded with the PIN diode to avoid the effect on mmWave antennas.



\begin{figure}[!t]
	\centering
	\includegraphics[width=70mm]{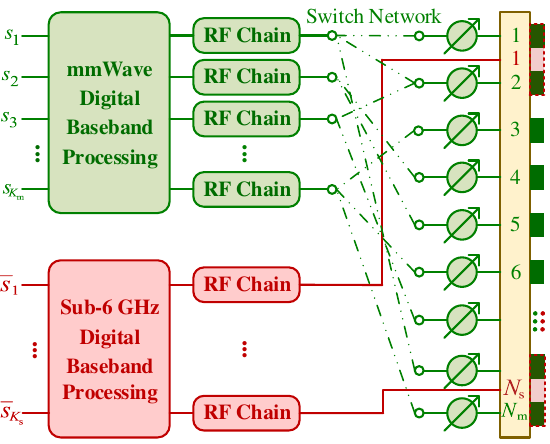}
	\caption{Illustration of the beamforming structure for the dual-band reconfigurable antenna array.}
	\label{fig_image2}
\end{figure}

\subsubsection{RAS Structure for the Sub-6G Band} To fully exploit the benefits of the DBRAA, we propose an RAS structure for the sub-6G band, as illustrated in Fig.~\ref{fig_AntennaSelection}. Note that a sub-6G antenna is formed by combining $2\times2$ mmWave antennas.  Consequently, there are $N_{\rm p} \triangleq (N_{\rm row} - 1) \times (N_{\rm col}-1)$ candidate sub-6G antennas. In addition, the spacing between adjacent  candidate sub-6G antennas is $\lambda_{\rm m}/2$ due to the half-wavelength-interval mmWave array and the reconfigurable formation of sub-6G antennas. Note that this spacing represents the nominal electrical separation between antenna centers rather than their absolute physical boundaries. Therefore, the RAS structure can support more  precise antenna position adjustment than the conventional antenna selection (CAS) structure.  The proposed RAS  structure dynamically selects $N_{\rm s}$ sub-6G antennas via a switch network from the $N_{\rm p}$ candidates for sub-6G signal transmission according to the CSI and system requirements. Typically, the number of antennas for a sub-6G antenna array is small  and the cost of the fully digital (FD) structure is affordable, where each of the $N_{\rm s}$ selected antennas is solely connected to one RF chain. Subsequently, a digital beamformer is adopted to allocate the $N_{\rm s}$ RF chains to $K_{\rm s}$ data streams, as illustrated in Fig.~\ref{fig_image2}. To support independent data transmission, we have $K_{\rm s}\le N_{\rm s}$.

\subsubsection{RHB Structure for the mmWave Band}
In the mmWave band, the number of antennas is usually very large. As a result, the full-digital structure is impractical and the more efficient hybrid beamforming structure is preferred. In this condition, we propose an RHB structure for the mmWave band. As illustrated in Fig.~\ref{fig_image2}, each of the $N_{\rm m}$ antennas is solely connected to one phase shifter. Then, the $N_{\rm m}$ phase shifters are connected to $N_{\rm RF}$ RF chains via a switch network. Subsequently, the $N_{\rm RF}$ RF chains are assigned to $K_{\rm m}$ data streams through the digital baseband processing. To ensure independent data transmission, we have $K_{\rm m}\le N_{\rm RF}$.  Note that the proposed RHB structure for mmWave systems differs from the dynamic hybrid beamforming structure (DHB) in \cite{TWC17PS,TWC24JX,TVT24WBW}. In the DHB structure, each antenna is solely connected to one RF chain.  However, in the  proposed RHB structure, each antenna can be dynamically connected to multiple  RF chains according to the CSI and system requirements. Compared to the DHB, the RHB retains similar  hardware complexity while increasing the design flexibility and improving the beamforming gain.



\begin{figure}[!t]
	\centering
	\includegraphics[width=77mm]{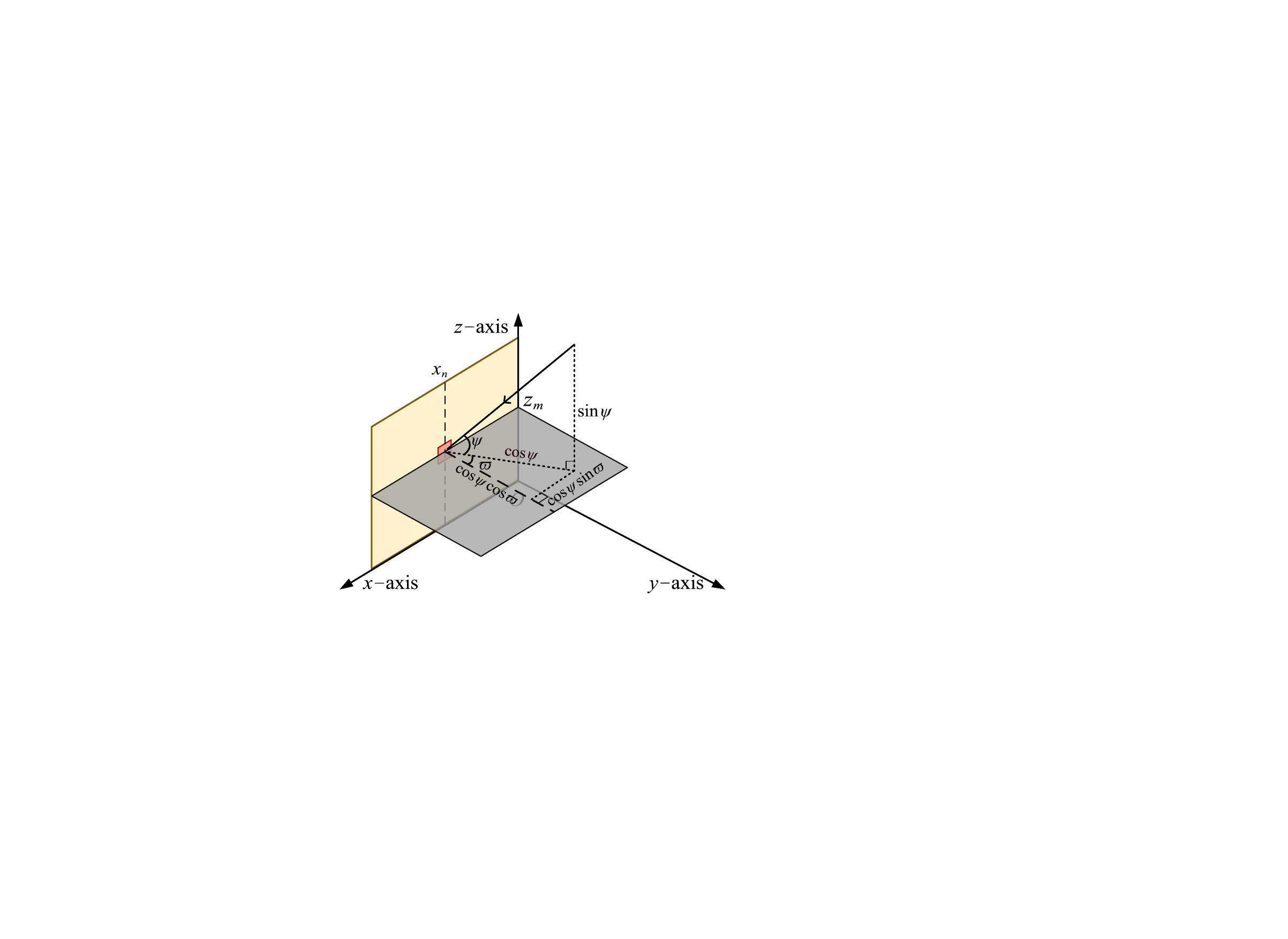}
	\caption{Illustration of the Cartesian coordinate system for the sub-6G channel.}
	\label{Coordinate}
\end{figure}


\subsection{Sub-6 GHz System Model}\label{SSM}

The received signals of the $K_{\rm s}$ sub-6G users can be expressed as 
\begin{align}\label{ReceivedSginal}
	\boldsymbol{y}_{\rm s} = \boldsymbol{H}_{\rm s}^{\rm H} \mathrm{diag}\{\boldsymbol{p}\}\boldsymbol{F}_{\rm s}\boldsymbol{s}_{\rm s} + \boldsymbol{n}_{\rm s},
\end{align}
where $\boldsymbol{H}_{\rm s}\in\mathbb{C}^{N_{\rm p}\times K_{\rm s}}$, $\boldsymbol{F}_{\rm s}\in\mathbb{C}^{N_{\rm p}\times K_{\rm s}}$, and $\boldsymbol{s}_{\rm s}\in\mathbb{C}^{K_{\rm s}}$ denote the sub-6G channel between ISAC BS and $K_{\rm s}$ users, the digital beamformer, and the data symbols of $K_{\rm s}$ users, respectively. In~\eqref{ReceivedSginal}, $\boldsymbol{p} \triangleq {\rm vec}\{\boldsymbol{P}\}$ and $\boldsymbol {P}\in\mathbb{Z}^{ (N_{\rm row}-1)\times(N_{\rm col}-1)}$ denotes the selection matrix.   $\boldsymbol{n}_{\rm s}\in\mathbb{C}^{K_{\rm s}}$ denotes the additive Gaussian noise, where $\boldsymbol{n}_{\rm s}\sim\mathcal{CN}(\boldsymbol{0},\sigma_{\rm s}^2\boldsymbol{I}_{K_{\rm s}})$. $\boldsymbol{h}_{{\rm s},k}\triangleq [\boldsymbol{H}_{\rm s}]_{:,k}$ denotes the channel between the ISAC BS and the $k$th sub-6G user, for $k = 1,2,\ldots,K_{\rm s}$.

Typically, the channel between the BS and a user is influenced by both the antenna positions and the propagation environment. Fig.~\ref{Coordinate} illustrates the Cartesian coordinate system employed to model these factors, where the location of the candidate  sub-6G  antenna at the $m$th row and  $n$th column is represented as $[x_n,0,z_m]$.
	
For a propagation path characterized by a physical azimuth angle  $\varpi$ and a physical elevation angle $\psi$, the corresponding channel steering vector is expressed as
\begin{align}\label{Original}
	\boldsymbol{\chi}(\psi,\varpi) =  e^{j2\pi\boldsymbol{z}\sin\psi/\lambda_{\rm s}}\otimes e^{j2\pi\boldsymbol{x}\cos\psi\sin\varpi/\lambda_{\rm s}},
\end{align}
where $\boldsymbol{z}\triangleq[z_1,\cdots,z_{N_{\rm row}-1}]^{\rm T}$ and  $\boldsymbol{x}\triangleq[x_1,\cdots,x_{N_{\rm col}-1}]^{\rm T}$ represent the spatial coordinate vectors along the $z$ and $x$ axes, respectively. To simplify the formulation, we introduce the surrogate azimuth angle  $\theta\triangleq\cos\psi\sin\varpi$ and the surrogate elevation angle $\phi\triangleq\sin\psi$. In addition, considering the configurations of the sub-6G antenna array in Section~\ref{Dual_bandReconAA}, we have $x_n = (n-1)\lambda_{\rm m}/2$ and $z_m = (m-1)\lambda_{\rm m}/2$.  Substituting these expressions into \eqref{Original}, the channel steering vector can be reformulated~as
\begin{align}\label{StBeta}
	&\boldsymbol{\beta}(\phi,\theta)\!=\!\boldsymbol{\gamma}(N_{\rm row}\!-\!1,\phi)\otimes\boldsymbol{\gamma}(N_{\rm col}\!-\!1,\theta),
\end{align}
with $\boldsymbol{\gamma}(\cdot)$ given by
\begin{align}
	&\boldsymbol{\gamma}(N,\theta) = \left[1,\cdots,e^{j\pi(n-1)\frac{\lambda_{\rm m}}{\lambda_{\rm s}}\theta},\cdots,e^{j\pi(N-1)\frac{\lambda_{\rm m}}{\lambda_{\rm s}}\theta}\right]^{\rm T}.
\end{align}
By superimposing the contributions from multiple propagation paths, $\boldsymbol{h}_{{\rm s},k}$ can be modeled as
\begin{align}\label{sub6GChannel}
	\boldsymbol{h}_{{\rm s},k} = \sqrt{\frac{1}{L_{{\rm s},k}}}\sum_{l=1}^{L_{{\rm s},k}} \zeta_{{\rm s}, k,l}\boldsymbol{\beta}(\phi_{{\rm s}, k,l},\theta_{{\rm s}, k,l}),
\end{align}
where $L_{{\rm s},k}$ denotes the number of channel paths between the ISAC BS and the $k$th sub-6G user. $\zeta_{{\rm s}, k,l}$, $\phi_{{\rm s}, k,l}$, and $\theta_{{\rm s}, k,l}$ are the channel gain, surrogate elevation angle, and surrogate azimuth angle of the $l$th path, respectively. 


\subsection{mmWave System Model}\label{MSM}
The received signals of the $K_{\rm m}$ mmWave users can be expressed as 
\begin{align}\label{ReceivedSginal2}
	\boldsymbol{y}_{\rm m} = \boldsymbol{H}_{\rm m}^{\rm H}\boldsymbol{F}_{\rm RF}\boldsymbol{F}_{\rm BB}\boldsymbol{s}_{\rm m} + \boldsymbol{n}_{\rm m},
\end{align}
where $\boldsymbol{H}_{\rm m}\in\mathbb{C}^{N_{\rm m}\times K_{\rm m}}$, $\boldsymbol{F}_{\rm RF}\in\mathbb{C}^{N_{\rm m}\times N_{\rm RF}}$, $\boldsymbol{F}_{\rm BB}\in\mathbb{C}^{N_{\rm RF}\times K_{\rm m}}$, and $\boldsymbol{s}_{\rm m}\in\mathbb{C}^{K_{\rm m}}$ denote the mmWave channel between ISAC BS and $K_{\rm m}$ users, the analog beamformer, the digital beamformer, and the data symbols of $K_{\rm m}$ users, respectively. $\boldsymbol{n}_{\rm m}\in\mathbb{C}^{K_{\rm m}}$ denotes the additive Gaussian noise, where $\boldsymbol{n}_{\rm m}\sim\mathcal{CN}(\boldsymbol{0},\sigma_{\rm m}^2\boldsymbol{I}_{K_{\rm m}})$. $\boldsymbol{h}_{{\rm m},k}\triangleq [\boldsymbol{H}_{\rm m}]_{:,k}$ denotes the channel between the ISAC BS and the $k$th mmWave user, for $k = 1,2,\ldots,K_{\rm m}$. Similar to \eqref{sub6GChannel}, $\boldsymbol{h}_{{\rm m},k}$ can be modeled as
\begin{align}
	&\boldsymbol{h}_{{\rm m},k}  = \sqrt{\frac{1}{L_{{\rm m},k}}}\!\sum_{l=1}^{L_{{\rm m},k}}\!\zeta_{{\rm m},k,l}\boldsymbol{\alpha}(N_{\rm row},\phi_{{\rm m},k,l})\otimes\boldsymbol{\alpha}(N_{\rm col},\theta_{{\rm m},k,l}),
\end{align}
where $L_{{\rm m},k}$ denotes  the number of channel paths between the ISAC BS and the $k$th mmWave user. $\zeta_{{\rm m},k,l}$, $\phi_{{\rm m},k,l}$, and $\theta_{{\rm m},k,l}$ represent the channel gain,  surrogate elevation angle, and surrogate azimuth angle of the $l$th path, respectively. $\boldsymbol{\alpha}(\cdot)$ denotes the mmWave channel steering vector, defined as 
\begin{align}
	\boldsymbol{\alpha}(N,\theta) = \left[1,e^{j\pi\theta},\ldots,e^{j\pi(N-1)\theta}\right]^{\rm T}.
\end{align}

\section{Problem Formulation}\label{ProblemFormulation}
In this section, we formulate a sub-6G and mmWave dual-band ISAC beamforming design problem, which aims at maximizing the mmWave  communication sum-rate subject to the constraints of the sub-6G communication QoS  and the sensing beamforming gain requirements. Then, the formulated dual-band ISAC beamforming design is further decoupled into the sub-6G beamforming design and the mmWave beamforming~design.

\subsection{Problem Formulation for Sub-6G Band}\label{SecProblemFormulation}


In the sub-6G  band, known for its robust penetration and extensive coverage, the ISAC BS aims to provide high-reliability communications for $K_{\rm s}$ sub-6G users. To ensure reliable transmission, the signal-to-interference-plus-noise ratio (SINR) of each user must exceed a predefined threshold $\Gamma_k$, i.e.,
\begin{align}\label{Sub6GSINR}
S_{{\rm s},k} \ge \Gamma_k,~k=1,2,\cdots,K_{\rm s}.
\end{align}
$S_{{\rm s},k}$ can be calculated as 
\begin{align}\label{SINR2}
	S_{{\rm s},k} = \frac{|\boldsymbol{h}_{{\rm s},k}^{\rm H}\mathrm{diag}\{\boldsymbol{p}\}\boldsymbol{f}_{{\rm s},k}|^2}{\sum_{i=1,i\neq k}^{K_{\rm s}}|\boldsymbol{h}_{{\rm s},k}^{\rm H}\mathrm{diag}\{\boldsymbol{p}\}\boldsymbol{f}_{{\rm s},i}|^2+ \sigma_{\rm s}^2  },
\end{align}
where $\boldsymbol{f}_{{\rm s},k}\triangleq [\boldsymbol{F}_{\rm s}]_{:,k}$. By neglecting inter-user interference, an upper bound for $\Gamma_{k}$ can be derived, which defines the maximum threshold beyond which the formulated problem has no feasible solution.  This bound serves as a reference point for iteratively adjusting $\Gamma_{k}$ to identify a suitable value that satisfies the system constraints.

In addition, the ISAC BS needs to sense the $T_{\rm s}$ directions for detecting potential targets. Therefore, the beamforming gain of the $t$th direction should also exceed a predefined threshold $\varUpsilon_{{\rm s},t}$, i.e.,
\begin{align}\label{Sub6GBG}
	G_{{\rm s},t} \ge \varUpsilon_{{\rm s},t},~t=1,\cdots,T_{\rm s}.
\end{align}
Denote the surrogate azimuth and elevation angles of the $t$th target as $\vartheta_{{\rm s},t}$  and $\varphi_{{\rm s},t}$, respectively. Then $G_{{\rm s},t}$ can be calculated as 
\begin{align}\label{BG2}
	G_{{\rm s},t} &= \mathbb{E}\{\|\boldsymbol{\beta}(\varphi_{{\rm s},t},\vartheta_{{\rm s},t})^{\rm H}\mathrm{diag}\{\boldsymbol{p}\}\boldsymbol{F}_{\rm s}\boldsymbol{s}_{\rm s}\|_2\}\nonumber \\
	&=\|\boldsymbol{F}_{\rm s}^{\rm H}\mathrm{diag}\{\boldsymbol{p}\}\boldsymbol{\beta}(\varphi_{{\rm s},t},\vartheta_{{\rm s},t})\|_2.
\end{align}
By applying Parseval's theorem to connect beamforming gains with transmit power~\cite{Tcom17SJH}, an upper bound for  $\varUpsilon_{{\rm s},t}$ can be derived, which defines the maximum threshold beyond which the formulated problem has no feasible solution.

Note that each entry in $\boldsymbol {P}$ can only be ``0" or ``1".  If one entry in the $\boldsymbol {P}$ is ``1", the corresponding sub-6G antenna is selected; otherwise, it is not selected. Therefore, $\boldsymbol{P}$ should satisfy 
\begin{align}\label{ASC}
	[\boldsymbol{P}]_{m,n}\! \in\! \{0,1\}, m = 1,\cdots,N_{\rm row}-1,~n = 1,\cdots,N_{\rm col}-1.
\end{align}

Note that only $N_{\rm s}$ sub-6G antennas are selected from $N_{\rm p}$ potential antennas. Therefore, the summation of all entries in $\boldsymbol {P}$ is $N_{\rm s}$, i.e.,
\begin{align}\label{ASTC}
	\boldsymbol{p}^{\rm T}\boldsymbol{1} = N_{\rm s}.
\end{align}

%

In the DBRAA, the spacing between adjacent candidate sub-6G antennas is $\lambda_{\rm m}/2$, which is much smaller than the sub-6G carrier wavelength due to the much higher carrier frequency of the mmWave band. For example, in a dual-band ISAC system operating at $5$ GHz and $30$ GHz, we have $\lambda_{\rm m}/2 = 0.5$ cm and $\lambda_{\rm s} = 6$ cm. If adjacent candidate sub-6G antennas are selected, strong coupling effects will be induced and system performance will be severely deteriorated. To address this, we stipulate that the distances between selected antennas must exceed a predefined threshold. Let $H_{\rm s}$ and $V_{\rm s}$ represent the minimum horizontal and vertical spacings between selected antennas, respectively. Let $J_{{\rm H},p}$ and $J_{{\rm V},p}$ denote the horizontal and vertical indices of the $p$th selected antenna, respectively. Then, we have 
\begin{align}\label{ASSC}
	|J_{{\rm H},p} - J_{{\rm H},q}|\ge H_{\rm s},~\mbox{and}~|J_{{\rm V},p} - J_{{\rm V},q}| \ge V_{\rm s},
\end{align}
for $p,q=1,2,\cdots,N_{\rm s}$ and $p\neq q$.
\subsection{Problem Formulation  for mmWave Band}
In the mmWave band, which is characterized by abundant spectrum resources and high data rates, the ISAC BS provides high-throughput communications for $K_{\rm m}$ mmWave users. Therefore, our objective is to maximize the sum-rate of these $K_{\rm m}$ mmWave users, which can be expressed as
\begin{align}\label{Rm}
	R_{\rm m} = \sum_{k=1}^{K_{\rm m}} \log_2(1+ S_{{\rm m},k}),
\end{align}
where $S_{{\rm m},k}$ denotes the SINR of the $k$th mmWave user and can be calculated as
\begin{align}
	S_{{\rm m},k} = \frac{|\boldsymbol{h}_{{\rm m},k}^{\rm H}\boldsymbol{F}_{\rm RF}[\boldsymbol{F}_{\rm BB}]_{:,k}|^2}{\sum_{i=1,i\neq k}^{K_{\rm m}}|\boldsymbol{h}_{{\rm m},k}^{\rm H}\boldsymbol{F}_{\rm RF}[\boldsymbol{F}_{\rm BB}]_{:,i}|^2+ \sigma_{\rm m}^2  }.
\end{align}
As shown in Fig.~\ref{fig_image2}, the analog beamformer is composed of the phase shifters and the switch network. Therefore, we have 
\begin{align}\label{APC}
  \boldsymbol{F}_{\rm RF} = {\rm diag}\{\boldsymbol{f}_{\rm P} \}\boldsymbol{S},
\end{align}
where $\boldsymbol{f}_{\rm P}\in\mathbb{C}^{N_{\rm m}}$ and $\boldsymbol{S}\in\mathbb{C}^{N_{\rm m}\times N_{\rm RF}}$ denote the phase shifters and switch network, respectively. Due to the constant modulus constraint of phase shifters, entries in $\boldsymbol{f}_{\rm P}$ should satisfy
\begin{align}\label{CMC}
\big|[\boldsymbol{f}_{\rm P}]_p\big| = 1, p =1,2,\cdots,N_{\rm m}.
\end{align}
In addition, due to the binary states of switches,  each entry in $\boldsymbol {S}$ can only be ``0" or ``1".  If one entry in  $\boldsymbol {S}$ is ``1",  the corresponding RF chain (column index) is connected to the corresponding antenna (row index); otherwise, the RF chain is not connected to the antenna. Therefore, we have 
\begin{align}\label{SC}
	[\boldsymbol{S}]_{p, u} \in {0,1},~p =1,2,\cdots,N_{\rm m},~u=1,2,\cdots,N_{\rm RF}.
\end{align}
Moreover,  the beamforming gain for $t$th mmWave target should also exceed a predefined threshold $\varUpsilon_{{\rm m}, t}$. Denote the surrogate azimuth and elevation angles of the $t$th mmWave target as $\vartheta_{{\rm m}, t}$ and $\varphi_{{\rm m}, t}$, respectively. Similar to \eqref{Sub6GBG} and \eqref{BG2}, we have
\begin{align}\label{MBGC}
	G_{{\rm m},t} \ge \varUpsilon_{{\rm m},t},
\end{align}
where 
\begin{align}
	G_{{\rm m},t}\! =\!  \big\|\boldsymbol{F}_{\rm BB}^{\rm H}\boldsymbol{F}_{\rm RF}^{\rm H}(\boldsymbol{\alpha}(N_{\rm row},\! \varphi_{{\rm m}, t})\! \otimes\!  \boldsymbol{\alpha}(N_{\rm col},\! \vartheta_{{\rm m}, t}))\big\|_2.
\end{align}

\subsection{Dual-Band Problem Formulation}
For the dual-band ISAC system, we aim at maximizing the sum-rate of $K_{\rm m}$ mmWave users subject to the constraints of the sub-6G communication SINR, sub-6G sensing beamforming gain requirements, mmWave sensing beamforming gain requirements, and the hardware structure. The dual-band ISAC beamforming design problem can be formulated as 
\begin{subequations}\label{TotalConstraint}
	\begin{align}
		&\max_{\boldsymbol{F}_{\rm s},\boldsymbol{P},\boldsymbol{f}_{\rm P},\boldsymbol{S},\boldsymbol{F}_{\rm BB}}~R_{\rm m}\\
		&~~~~~~~\mathrm{s.t.}\!~~~~~~~~\eqref{Sub6GSINR},\!~\eqref{Sub6GBG},\!~\eqref{ASC},\!~\eqref{ASTC},\!~\eqref{ASSC}, \label{Sub6GCon}\\
		&~~~~~~~~~~~~~~~~~~\eqref{APC},\!~\eqref{CMC},\!~\eqref{SC},\!~\eqref{MBGC} \label{mmWaveCon},\\
		&~~~~~~~~~~~~~~~~~~\|\mathrm{diag}\{\boldsymbol{p}\}\boldsymbol{F}_{\rm s}\|_{\rm F}^2 \le P_{\rm s},\label{Sub6GP}\\
		&~~~~~~~~~~~~~~~~~~\|\boldsymbol{F}_{\rm RF}\boldsymbol{F}_{\rm BB}\|_{\rm F}^2\le P_{\rm m},\label{mmWaveP}\\
		&~~~~~~~~~~~~~~~~~~ P_{\rm s} + P_{\rm m}\le P_{\rm t}, \label{total}
	\end{align} 
\end{subequations}
where the sub-6G system constraints, mmWave system constraints, sub-6G power  constraint, mmWave power constraint,  and total power constraint are stated in
\eqref{Sub6GCon}, \eqref{mmWaveCon}, \eqref{Sub6GP}, \eqref{mmWaveP}, and \eqref{total}, respectively. $P_{\rm s}$ in \eqref{Sub6GP}, $P_{\rm m}$ in \eqref{mmWaveP}, and $P_{\rm t}$ in \eqref{total} denote the maximum sub-6G transmit power, the maximum mmWave transmit power, and the maximum total transmit power, respectively.

\textbf{Remark 4:} In \eqref{TotalConstraint}, we aim to maximize the mmWave multiuser sum-rate to achieve high throughput in the mmWave band while ensuring the sub-6G SINR performance to guarantee high reliability. This approach is motivated by the distinct characteristics and requirements of each frequency band. The SINR-based objective in the sub-6G band ensures communication reliability, which is critical for low-latency applications. However, it may result in suboptimal bandwidth utilization compared to throughput-maximizing objectives. Conversely, maximizing the sum-rate in the mmWave band ensures efficient spectrum usage and high throughput but may not adequately address reliability challenges in large-scale deployments. In this context, further refinement and adaptation of the proposed approaches can be conducted to meet the demands of practical system deployments.

From \eqref{TotalConstraint}, the sub-6G and mmWave systems have independent variables and constraints but are coupled through their respective transmit powers, $P_{\rm m}$ and $P_{\rm s}$, as defined in \eqref{total}. In the mmWave band, an increase in the transmit power $P_{\rm m}$ leads to a higher objective value $R_{\rm m}$. In addition, according to \eqref{total}, reducing the sub-6G transmit power allows for a higher mmWave transmit power. Consequently, decreasing the sub-6G transmit power improves the mmWave sum-rate.  Based on the above discussions, to solve \eqref{TotalConstraint}, we first design the sub-6G beamforming by minimizing the transmit power required to meet the SINR constraints of sub-6G users. Subsequently, we optimize the mmWave beamforming to maximize the mmWave sum-rate using the remaining available power. The sub-6G beamforming design can be formulated as

\begin{align}\label{TotalConstraintS1}
		(\mbox{P1})~~~~&\min_{\boldsymbol{F}_{\rm s},\boldsymbol{P}}~ \|\mathrm{diag}\{\boldsymbol{p}\}\boldsymbol{F}_{\rm s}\|_{\rm F}^2\nonumber\\
		&~\mathrm{s.t.}\!~~\eqref{Sub6GSINR},\!~\eqref{Sub6GBG},\!~\eqref{ASC},\!~\eqref{ASTC},\!~\eqref{ASSC}.
\end{align} 
Let $\overline{P}_{\rm s}$ be the transmit power corresponding to the solution of \eqref{TotalConstraintS1}. The power available for the mmWave system is $P_{\rm m} = P_{\rm t} - \overline{P}_{\rm s}$. Then, the mmWave beamforming design can be formulated as 

\begin{subequations}\label{TotalConstraintS2}
	\begin{align}
		(\mbox{P2})~~~~&\max_{\boldsymbol{f}_{\rm P},\boldsymbol{S},\boldsymbol{F}_{\rm BB}}~ R_{\rm m}\\
		&~~~\mathrm{s.t.}\!~~~~~~\eqref{APC},\!~\eqref{CMC},\!~\eqref{SC},\!~\eqref{MBGC},\\
		&~~~~~~~~~~~~\|\boldsymbol{F}_{\rm RF}\boldsymbol{F}_{\rm BB}\|_{\rm F}^2\le P_{\rm m}. \label{mmwavepower}
	\end{align} 
\end{subequations}
Note that we design the mmWave and sub-6G beamforming while considering their coupled power budget in \eqref{TotalConstraint}. Nevertheless, alternative system design considerations, such as dual-band performance constraints under QoS requirements~\cite{TVT20CM}, sum-rate maximization with fairness guarantees~\cite{TVT23LJW}, and joint resource allocation strategies~\cite{TWC17SO}, could be explored in future extensions. In the following, we design the beamforming for sub-6G systems  by solving \eqref{TotalConstraintS1} in Section~\ref{SSD}, and the beamforming for mmWave systems by solving \eqref{TotalConstraintS2} in Section~\ref{MSD}.

\section{Sub-6G Beamforming Design}\label{SSD}
In this section, we design the sub-6G beamforming by solving (P1) in \eqref{TotalConstraintS1}, where an FS-JBAS algorithm is developed.

\subsection{Subproblem of (P1)}\label{intialFSJBAS}
Due to the complex and nonconvex constraints in \eqref{Sub6GSINR}, \eqref{Sub6GBG}, \eqref{ASC}, and \eqref{ASSC}, (P1) is challenging to solve. To address this, we consider a subproblem of (P1). The solution to this subproblem is used as an initial value to aid  the convergence of the FS-JBAS algorithm.

The most challenging constraint is the antenna spacing constraint in \eqref{ASSC}. Therefore, in the considered subproblem, we ensure that the candidate antennas satisfy the antenna spacing constraint. We define a matrix $\boldsymbol{C}\in\mathbb{C}^{(N_{\rm row}-1)\times(N_{\rm col}-1)}$ to represent the candidate antennas as
\begin{align}
	[\boldsymbol{C}]_{m,n}\!=\!\begin{cases}
		1,~m \!=\!bV_{\rm s}\! + \!1, ~n \!=\! dH_{\rm s}\! +\! 1, b,d\!\in\!\mathbb{Z},\\
		0,~\mathrm{others}.
	\end{cases}
\end{align}
If an entry in $\boldsymbol{C}$ is "1", the corresponding sub-6G antenna can be selected; otherwise, it cannot be selected. We also define a selection matrix $\boldsymbol{Q}\in\mathbb{C}^{M_{\rm row}\times M_{\rm col}}$ to represent the selection states of the candidate antennas represented by $\boldsymbol{C}$, where $M_{\rm row} = \lceil \frac{N_{\rm row} - 1}{V_{\rm s}} \rceil$ and $M_{\rm col} = \lceil \frac{N_{\rm col} - 1}{H_{\rm s}} \rceil$ denote the number of rows and columns of candidate antennas, respectively. If an entry in the $b$th row and $d$th column of $\boldsymbol{Q}$ is non-zero, for $b = 1, 2, \cdots, M_{\rm row}$ and $d = 1, 2, \cdots, M_{\rm col}$, it means that the sub-6G antenna on the $((b-1)V_{\rm s} + 1)$th row and the $((d-1)H_{\rm s} + 1)$th column of the array is selected.

By selecting the antennas from the candidate antennas represented by $\boldsymbol{C}$, we can convert (P1) in \eqref{TotalConstraintS1} to
\begin{subequations}\label{SubProblemP1S}
\begin{align}
	(\mbox{P1-S})~~~~&\min_{\boldsymbol{F}_{\rm s},\boldsymbol{P}}~ \|\mathrm{diag}\{\boldsymbol{p}\}\boldsymbol{F}_{\rm s}\|_{\rm F}^2\label{P1S-1}\\
	&~\mathrm{s.t.}\!~~~\eqref{Sub6GSINR},\!~\eqref{Sub6GBG},\!~\eqref{ASTC},\label{P1S-2}\\
	&~~~~~~~\boldsymbol{P} = \mathcal{T}(\boldsymbol{Q}), \label{P1S-3}\\
	&~~~~~~~[\boldsymbol{Q}]_{b,d}\in\{0,1\}. \label{P1S-4}
\end{align} 
\end{subequations}
The constraint in \eqref{P1S-3} represents  the transformation from  $\boldsymbol{Q}$ to $\boldsymbol{P}$, which can be expressed as
\begin{align}\label{transformation}
	[\boldsymbol{P}]_{m,n}\!=\!\begin{cases}
		[\boldsymbol{Q}]_{b,d},\begin{array}{l}
			m  = bV_{\rm s}  + 1,~n  =  dH_{\rm s}  + 1,\\
			b =1,\cdots,M_{\rm row},~d =1,\cdots,M_{\rm col},\\
		\end{array}\\
		0,~~~~~~~\mathrm{others}.
	\end{cases}
\end{align}
Note that \eqref{transformation} represents linear equations. Therefore, \eqref{P1S-3} is a convex constraint. In addition, the constraint in \eqref{ASSC} is omitted because the horizontal and vertical spacings between two arbitrary non-zero entries in $\boldsymbol{C}$ are no less than $H_{\rm s}$ and $V_{\rm s}$, respectively. In \eqref{SubProblemP1S}, since using either $\boldsymbol{P}$ or $\boldsymbol{Q}$ as the optimization variable yields equivalent formulations, for consistency with other sections, we use $\boldsymbol{P}$ as the optimization variable. We then propose an alternating beamforming and antenna selection algorithm (ASBS) to solve \eqref{SubProblemP1S}.

First, we address the integer 0-1 constraint in $\eqref{P1S-4}$ and reformulate it as~\cite{liu2023joint}
\begin{subequations}\label{binaryconstraint}
\begin{align}
	&\min_{\boldsymbol{q}} \boldsymbol{q}^{\rm T}(\boldsymbol{1}-\boldsymbol{q}) \label{cobj} \\
	&~\mathrm{s.t.}~ \boldsymbol{q} = \mathrm{vec}\{\boldsymbol{Q}\},~0\le[\boldsymbol{Q}]_{b,d}\le1.\label{cconst}
\end{align}
\end{subequations}
Based on \eqref{binaryconstraint}, we introduce a  weighting coefficient $\mu $, and take the weighted sum of \eqref{P1S-1} and \eqref{cobj} as the objective. Then, \eqref{SubProblemP1S} can be converted to
\begin{align}\label{SubProblemP2S}
		&\min_{\boldsymbol{F}_{\rm s},\boldsymbol{P}}~ \|\mathrm{diag}\{\boldsymbol{p}\}\boldsymbol{F}_{\rm s}\|_{\rm F}^2 + \mu 
		\boldsymbol{q}^{\rm T}(\boldsymbol{1}-\boldsymbol{q}) \nonumber\\
		&~\mathrm{s.t.}~~~\eqref{Sub6GSINR},~\eqref{Sub6GBG},~\eqref{ASTC},~\eqref{P1S-3},~\eqref{cconst}.
\end{align} 
Note that \eqref{SubProblemP2S} involves the digital beamformer, $\boldsymbol{F}_{\rm s}$, and the antenna selection matrix, $\boldsymbol{P}$. Following the existing works~\cite{JSTSP16YXH}, we employ the alternating minimization method to solve it.

\subsubsection{Initialization}\label{Ini}
To ensure high beamforming gains for communication users and sensing targets, we initialize $\boldsymbol{F}_{\rm s}$ as 
\begin{align}
[\boldsymbol{F}_{\rm s}]_{:,k} = \kappa\bigg(\boldsymbol{h}_{{\rm s},k} + \sum_{t=1}^{T}\boldsymbol{\beta}(\boldsymbol{P},\varphi_{{\rm s},t},\vartheta_{{\rm s},t})\bigg),
\end{align}
where $\kappa$ is the scaling factor. In practice, $\kappa$ can be set to a large number, such as 10, to ensure that the optimization in \eqref{SubProblemP2S} has feasible solutions initially. In addition,  to satisfy the antenna selection constraint in \eqref{ASTC}, we initialize $\boldsymbol{P}$ as 
\begin{align}\label{initializeFsP}
  [\boldsymbol{P}]_{m,n} = N_{\rm s}/(M_{\rm row}M_{\rm col}).
\end{align}


\subsubsection{Optimization of $\boldsymbol{F}_{\rm s}$}\label{OPTFs}
Denote the solution of $\boldsymbol{P}$ in the previous iteration as $\boldsymbol{\overline{P}}$. Then, \eqref{SubProblemP2S} can be reformulated as 
\begin{align}\label{subproblemFs}
	&\min_{\boldsymbol{F}_{\rm s}}~ \|\mathrm{diag}\{\boldsymbol{\overline{p}}\}\boldsymbol{F}_{\rm s}\|_{\rm F}^2 \nonumber\\
	&~\mathrm{s.t.}~~~\eqref{Sub6GSINR},~\eqref{Sub6GBG},
\end{align}
where $\boldsymbol{\overline{p}} = \mathrm{vec}\{\boldsymbol{\overline{P}}\}$. Note that \eqref{subproblemFs} is a nonlinear programming problem with convex objective and nonconvex constraints. In the following, we will address these nonconvex constraints to develop a tractable optimization problem.

First, we focus on the nonconvex constraint in \eqref{Sub6GSINR}. By exploiting the phase rotation property of complex values, the constraint in \eqref{Sub6GSINR} can be equivalently converted to a second-order cone constraint~\cite{SPM14BE}, expressed as 
\begin{align}\label{socc}
\left\|\left[\begin{array}{c}
	\boldsymbol{F}_{\rm s}^{\rm H}\mathrm{diag}\{\boldsymbol{\overline{p}}\}\boldsymbol{h}_{{\rm s},k}\\
	\sigma_{\rm s}\\
\end{array}\right]\right\|_2\le\sqrt{1+\frac{1}{\Gamma_k}}\mathcal{R}\{\boldsymbol{f}_{{\rm s},k}^{\rm H}\mathrm{diag}\big\{\boldsymbol{\overline{p}}\big\}\boldsymbol{h}_{{\rm s},k}\}.
\end{align}

Next, we focus on the nonconvex constraint in \eqref{Sub6GBG}. Note that  a quadratic convex function $\boldsymbol{x}^{\rm H}\boldsymbol{A}\boldsymbol{x}$ can be lower-bounded by its first-order Taylor expansion, i.e.,  
\begin{align}\label{lowerbound}
	\boldsymbol{x}^{\rm H}\boldsymbol{A}\boldsymbol{x}&\ge \boldsymbol{\overline{x}}^{\rm H}\boldsymbol{A}\boldsymbol{\overline{x}} + 2\mathcal{R}\{\boldsymbol{\overline{x}}^{\rm H}\boldsymbol{A}(\boldsymbol{x}-\boldsymbol{\overline{x}})\} \nonumber\\
	&=2\mathcal{R}\{\boldsymbol{\overline{x}}^{\rm H}\boldsymbol{A}\boldsymbol{x}\} - \boldsymbol{\overline{x}}^{\rm H}\boldsymbol{A}\boldsymbol{\overline{x}},
\end{align}
where $\boldsymbol{\overline{x}}$ denotes given values of $\boldsymbol{x}$. Then, $G_{{\rm s},t}$ in \eqref{BG2} can be lower-bounded by
\begin{align}\label{lowerbound_BG}
	G_{{\rm s},t} &= \|\boldsymbol{F}_{\rm s}^{\rm H}\mathrm{diag}\{\boldsymbol{\overline{p}}\}\boldsymbol{\beta}(\varphi_{{\rm s},t},\vartheta_{{\rm s},t})\|_2 \nonumber\\
	&= \boldsymbol{t}^{\rm H}\boldsymbol{\overline{\Phi}}\boldsymbol{t} \nonumber \\
	&\ge2\mathcal{R}\big\{\boldsymbol{\overline{t}}^{\rm H}\boldsymbol{\overline{\Phi}}\boldsymbol{t}\big\} - \boldsymbol{\overline{t}}^{\rm H}\boldsymbol{\overline{\Phi}}\boldsymbol{\overline{t}},
\end{align}
where $\boldsymbol{t}\triangleq \mathrm{vec}\{\boldsymbol{F}_{\rm s}\}$, $\overline{\boldsymbol{t}}\triangleq \mathrm{vec}\{\boldsymbol{\overline{F}}_{\rm s}\}$ denotes the solution of $\boldsymbol{t}$ in the $(i-1)$th iteration, $\boldsymbol{\overline{\Phi}}\triangleq \boldsymbol{I}_{ K_{\rm s}}\otimes\boldsymbol{\Phi}$, and $\boldsymbol{\Phi}\triangleq \mathrm{diag}\{\boldsymbol{\overline{p}}\}\boldsymbol{\beta}(\varphi_{{\rm s},t},\vartheta_{{\rm s},t})\boldsymbol{\beta}(\varphi_t,\vartheta_t)^{\rm H}\mathrm{diag}\{\boldsymbol{\overline{p}}\}$. Based on \eqref{lowerbound_BG}, the constraint in \eqref{Sub6GBG} can be relaxed as 
\begin{align}\label{relax_BG}
	2\mathcal{R}\big\{\boldsymbol{\overline{t}}^{\rm H}\boldsymbol{\overline{\Phi}}\boldsymbol{t}\big\} - \boldsymbol{\overline{t}}^{\rm H}\boldsymbol{\overline{\Phi}}\boldsymbol{\overline{t}}\ge \Upsilon_{{\rm s},t}.
\end{align}

With \eqref{socc} and \eqref{relax_BG}, \eqref{subproblemFs} can be converted to 
\begin{align}\label{subproblemFs2}
	&\min_{\boldsymbol{F}_{\rm s}}~ \|\mathrm{diag}\{\boldsymbol{\overline{p}}\}\boldsymbol{F}_{\rm s}\|_{\rm F}^2 \nonumber\\
	&~\mathrm{s.t.}~~~\eqref{socc},~\eqref{relax_BG},
\end{align}
which is a second-order cone programming (SOCP) problem and therefore can be effectively solved via convex optimization.

\subsubsection{Optimization of $\boldsymbol{p}$} Denote the solution of $\boldsymbol{F}_{\rm s}$ in the previous iteration as $\boldsymbol{\overline{F}}_{\rm s}$. Then, \eqref{SubProblemP2S} can be converted to 
\begin{align}\label{SubProblemFs}
	&\min_{\boldsymbol{P}}~ \|\mathrm{diag}\{\boldsymbol{p}\}\boldsymbol{\overline{F}}_{\rm s}\|_{\rm F}^2 + \mu 
	\boldsymbol{q}^{\rm T}(\boldsymbol{1}-\boldsymbol{q}) \nonumber\\
	&~\mathrm{s.t.}~~~\eqref{Sub6GSINR},~\eqref{Sub6GBG},~\eqref{ASTC},~\eqref{P1S-3},~\eqref{cconst}.
\end{align}
Similar to \eqref{socc} and \eqref{relax_BG}, \eqref{Sub6GSINR} and \eqref{Sub6GBG} can be respectively converted to 
\begin{subequations}
\begin{align}
		&\left\|\left[\begin{array}{c}
			\boldsymbol{\overline{F}}_{\rm s}^{\rm H}\mathrm{diag}\{\boldsymbol{h}_{{\rm s},k}\}\boldsymbol{p}\\
			\sigma_{\rm s}\\
		\end{array}\right]\right\|_2\le\sqrt{1+\frac{1}{\Gamma_k}}\mathcal{R}\{\boldsymbol{\overline{f}}_{{\rm s},k}^{\rm H}\mathrm{diag}\big\{\boldsymbol{h}_{{\rm s},k}\big\}\boldsymbol{p}\}\label{socc2}\\
		& 	2\mathcal{R}\big\{\boldsymbol{\overline{p}}^{\rm T}\boldsymbol{U}\boldsymbol{p}\big\} - \boldsymbol{\overline{p}}^{\rm T}\boldsymbol{U}\boldsymbol{\overline{p}}\ge \Upsilon_{{\rm s},t}\label{relax_BG2},
\end{align}
\end{subequations}
where $\boldsymbol{U}\triangleq \mathrm{diag}\{\boldsymbol{\beta}(\varphi_t,\vartheta_t)^{\rm H}\}\boldsymbol{\overline{F}}_{\rm s}\boldsymbol{\overline{F}}_{\rm s}^{\rm H}\mathrm{diag}\{\boldsymbol{\beta}(\varphi_t,\vartheta_t)\}$, and $\boldsymbol{\overline{f}}_{{\rm s},k}\triangleq [\boldsymbol{\overline{F}}_{\rm s}]_{:,k}$. Then, \eqref{SubProblemFs} can be converted to 
\begin{align}\label{SubProblemFs2}
	&\min_{\boldsymbol{P}}~ \|\mathrm{diag}\{\boldsymbol{p}\}\boldsymbol{\overline{F}}_{\rm s}\|_{\rm F}^2 + \mu 
	\boldsymbol{q}^{\rm T}(\boldsymbol{1}-\boldsymbol{q}) \nonumber\\
	&~\mathrm{s.t.}~~~\eqref{ASTC},~\eqref{P1S-3},~\eqref{cconst},~\eqref{socc2},~\eqref{relax_BG2}.
\end{align}
Note that \eqref{SubProblemFs2} is an optimization problem with convex constraints and nonconvex objective due to the minimization of the concave function. To address this issue, we employ the majorization-minimization (MM) method to convert the  objective in \eqref{SubProblemFs2} into a tractable one. The MM method aims at finding an upper-bound surrogate function of the original function around a given point. From \eqref{lowerbound}, we have 
\begin{align}
	\boldsymbol{q}^{\rm T}(\boldsymbol{1}-\boldsymbol{q})\le\boldsymbol{q}^{\rm T}\boldsymbol{1} - 2\mathcal{R}\{\boldsymbol{\overline{q}}^{\rm T}\boldsymbol{q}\} + \boldsymbol{\overline{q}}^{\rm T}\boldsymbol{\overline{q}},
\end{align}
where $\overline{\boldsymbol{q}}$ denotes the solution of $\boldsymbol{q}$ in the previous iteration. Based on the relationship between $\boldsymbol{P}$ and $\boldsymbol{Q}$ in \eqref{transformation}, we can easily obtain $\overline{\boldsymbol{q}}$.  Therefore, the surrogate function of the objective in \eqref{SubProblemFs2} can be expressed as 
\begin{align}\label{surrogatefunction}
	\|\mathrm{diag}\{\boldsymbol{p}\}\boldsymbol{\overline{F}}_{\rm s}\|_{\rm F}^2 - 2\mu \mathcal{R}\{\boldsymbol{\overline{q}}^{\rm T}\boldsymbol{q}\},
\end{align}
where we omit the constant terms $\boldsymbol{q}^{\rm T}\boldsymbol{1}$ and $\boldsymbol{\overline{q}}^{\rm T}\boldsymbol{\overline{q}}$. Then, we can convert \eqref{SubProblemFs2} to
\begin{align}\label{SubProblemFs3}
	&\min_{\boldsymbol{P}}~~ \eqref{surrogatefunction} \nonumber\\
	&~\mathrm{s.t.}~~~\eqref{ASTC},~\eqref{P1S-3},~\eqref{cconst},~\eqref{socc2},~\eqref{relax_BG2},
\end{align}
which is also an SOCP problem and can be effectively solved.

\begin{algorithm}[!t]
	\caption{Alternating Beamforming and Antenna Selection (ABAS) Algorithm}
	\label{alg_JBAS}
	\begin{algorithmic}[1]
		\STATE \textbf{Input:} $N_{\rm row}$, $N_{\rm col}$, $N_{\rm s}$, $H_{\rm s}$, $V_{\rm s}$, $\boldsymbol{H}_{\rm s}$, $\Gamma_k$, $\Upsilon_{{\rm s},t}$, $\varphi_t$, and $\vartheta_t$.
		\STATE \textbf{Initialization:} Initialize $\boldsymbol{F}_{\rm s}$ and $\boldsymbol{P}$ via \eqref{initializeFsP}, $i\leftarrow 0$.
		\WHILE{$i\le I_{\rm A}$}
		\STATE Obtain $\overline{\boldsymbol{F}}_{\rm s}$  via \eqref{subproblemFs2}.
		\STATE Obtain $\overline{\boldsymbol{P}}$  via \eqref{SubProblemFs3}.
		\STATE $i\leftarrow i+1$.
		\ENDWHILE
		\STATE Obtain $\widehat{\boldsymbol{P}}$ via \eqref{FinalP}.
		\STATE Obtain $\widehat{\boldsymbol{F}}_{\rm s}$ via \eqref{subproblemFs}.
		\STATE \textbf{Output:} $\widehat{\boldsymbol{P}}$ and $\widehat{\boldsymbol{F}}_{\rm s}$.
	\end{algorithmic}
\end{algorithm}

\subsubsection{Summary of ABAS Algorithm}\label{Summary} 
We alternately  optimize $\boldsymbol{F}_{\rm s}$ and $\boldsymbol{P}$ until the maximum number of iterations $I_{\rm A}$ is reached. The solutions of $\boldsymbol{F}_{\rm s}$ and $\boldsymbol{P}$ are denoted as $\boldsymbol{\widetilde{F}}_{\rm s}$ and $\boldsymbol{\widetilde{P}}$, respectively. In \eqref{SubProblemP2S}, the values of $\boldsymbol{P}$ are relaxed as continuous between zero and one. However, the antenna selection matrix $\boldsymbol{P}$ should take binary values of either zero or one. To avoid computational deviations, we quantize $\boldsymbol{\widetilde{P}}$ as
\begin{align}\label{FinalP}
	[\boldsymbol{\widehat{P}}]_{m,n} = \mathrm{round}\big([\boldsymbol{\widetilde{P}}]_{m,n}\big),
\end{align}
where $\mathrm{round}(\cdot)$ denotes the rounding operation. Then, we solve \eqref{subproblemFs} to update $\boldsymbol{\widetilde{F}}_{\rm s}$ as $\boldsymbol{\widehat{F}}_{\rm s}$.

Finally, we summarize the alternating beamforming and antenna selection algorithm in \textbf{ Algorithm~\ref{alg_JBAS}}.

\subsection{Detailed Implementation of the FS-JBAS Algorithm}\label{DIFSJBAS}

Based on Section \ref{intialFSJBAS}, the initial value of $\boldsymbol{P}$ is given by $\underline{\boldsymbol{P}}^{(0)} = \widehat{\boldsymbol{P}}$. From \eqref{ASTC}, $\underline{\boldsymbol{P}}^{(0)}$ has $N_{\rm s}$ non-zero entries, which correspond to $N_{\rm s}$ selected sub-6G antennas. We denote the row and column indices of these antennas  as  $\boldsymbol{x}^{(0)}\in \mathbb{Z}^{N_{\rm s}}$ and $\boldsymbol{y}^{(0)}\in \mathbb{Z}^{N_{\rm s}}$, respectively. We also define a transformation function $\mathcal{F}(\cdot)$ that maps the indices  $\boldsymbol{x}^{(0)}$ and $\boldsymbol{y}^{(0)}$ to the selection matrix, which can be expressed as 
\begin{align}\label{transformation2}
	[\underline{\boldsymbol{P}}^{(0)}]_{m,n} = \begin{cases}
		1,~m = [\boldsymbol{x}^{(0)}]_p,~n=[\boldsymbol{y}^{(0)}]_p,~p=1,\cdots,N_{\rm s},\\
		0,~\mathrm{others}.
	\end{cases}
\end{align}
We then propose a fast search algorithm that iteratively updates the indices of each selected antenna in  $\underline{\boldsymbol{P}}^{(0)}$.

\begin{figure*}[t]
	\centering
	\includegraphics[width=185mm]{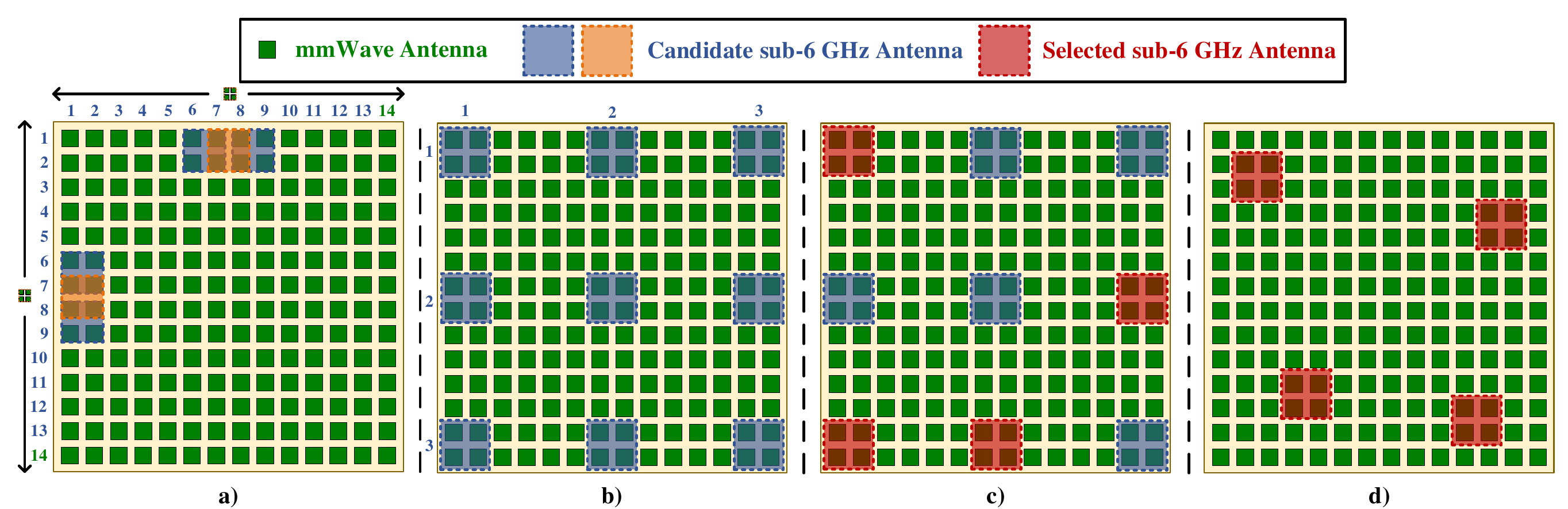}
	\caption{Illustration of the design example for sub-6G systems: a) Candidate sub-6G antennas of the DBRAA; b) Candidate sub-6G antennas of (P1-S); c) Selected sub-6G antennas by solving (P1-S); d) Selected sub-6G antennas by solving (P1). }
	\label{fig_DesignExample}
		\vspace{-0.5cm}
\end{figure*}
\begin{algorithm}[!t]
	\caption{Fast Search-based Joint Beamforming and Antenna Selection  (FS-JBAS) Algorithm }
	\label{alg_FSJBAS}
	\begin{algorithmic}[1]
		\STATE \textbf{Input:} $N_{\rm row}$, $N_{\rm col}$, $N_{\rm s}$, $H_{\rm s}$, $V_{\rm s}$, $\boldsymbol{H}_{\rm s}$, $\Gamma_k$, $\Upsilon_{{\rm s},t}$, $\varphi_t$, and $\vartheta_t$.
		\STATE Obtain $\boldsymbol{x}^{(0)}$, $\boldsymbol{y}^{(0)}$ and $\underline{\boldsymbol{P}}^{(0)}$ via \textbf{Algorithm~\ref{alg_JBAS}}.~$i\leftarrow 0$.
		\WHILE{stop conditions are not satisfied}
		\STATE Obtain $\overline{p}$  via \eqref{currentopta}.
		\STATE Obtain $\mathcal{S}_{i}$  via \eqref{candidateset}.
		\FOR{$s=1,2,\cdots,|\mathcal{S}_{i}|$}
		\STATE Obtain $\overline{\boldsymbol{x}}^{(s)}$ and $\overline{\boldsymbol{y}}^{(s)}$ via \eqref{indicesofxys}.
		\STATE Obtain $\overline{\overline{\boldsymbol{P}}}^{(s)}$ via \eqref{powerps}.
		\STATE Obtain $\widetilde{P}_{\rm s}^{(s)}$ via \eqref{designexample}.
		\ENDFOR
		\STATE Obtain $s^{*}$ via \eqref{minimumpowerindex}.
		\STATE Obtain $\boldsymbol{x}^{(i)}$ and  $\boldsymbol{y}^{(i)}$  via \eqref{xiyi}.
		\STATE Obtain $ \widehat{P}_{\rm s}^{(i)} $ via \eqref{ithminimum1}.
		\STATE $i\leftarrow i+1$.
		\ENDWHILE
		\STATE Obtain the current iteration number $\overline{i}$.
		\STATE $\overline{P}_{\rm s}\leftarrow\widehat{P}_{\rm s}^{(\overline{i})}$, $\overline{\boldsymbol{x}}\leftarrow\boldsymbol{x}^{(\overline{i})}$, and $\overline{\boldsymbol{y}}\leftarrow\boldsymbol{y}^{(\overline{i})}$.
		\STATE \textbf{Output:} $\overline{P}_{\rm s}$, $\overline{\boldsymbol{x}}$, and $\overline{\boldsymbol{y}}$.
	\end{algorithmic}
\end{algorithm}

In the $i$th iteration, for $i\ge 1$, the order of the antenna to be updated is 
\begin{align}\label{currentopta}
	\overline{p} = \mathrm{mod}(i-1,N_{\rm s}) + 1.
\end{align}
The row and column indices of this antenna in the previous iteration are $\big[\boldsymbol{x}^{(i-1)}\big]_{\overline{p}}$ and $\big[\boldsymbol{y}^{(i-1)}\big]_{\overline{p}}$, respectively. Then, we test the candidate antennas to optimize the position of the $\overline{p}$th antenna. Due to the antenna spacing constraint in \eqref{ASSC}, not all the candidate antennas can be selected as the $\overline{p}$th antenna. We express the set of row and column indices for all feasible candidate antennas~as
\begin{align}\label{candidateset}
	\mathcal{S}_{i} = \left\{[x,y]\bigg|\!\begin{array}{c}
		\big|x-[\boldsymbol{x}^{(i-1)}]_p\big|\!\ge\!V_{\rm s},\big|y-[\boldsymbol{y}^{(i-1)}]_p\big|\!\ge\!H_{\rm s},\\
		x\!=\!1,\!\cdots\!,N_{\rm row}\!-\!1,~y=1,\!\cdots\!,N_{\rm col}\!-\!1,\\
		p=1,2,\cdots,N_{\rm s},~p\neq\overline{p}\\
	\end{array}\right\},
\end{align}
where the number of elements in $\mathcal{S}_{i}$ can be expressed as its cardinality  $|\mathcal{S}_{i}|$.

Then, we sequentially test each element in $\mathcal{S}_{i}$ as follows. When testing the $s$th element, the row and column indices of the tested antenna can be expressed as $\big[\mathcal{S}_i\{s\}\big]_1$ and $\big[\mathcal{S}_i\{s\}\big]_2$, respectively. Then, the row and column indices of the $N_{\rm s}$ antennas are expressed as 
$\overline{\boldsymbol{x}}^{(s)}$ and $\overline{\boldsymbol{y}}^{(s)}$, respectively, where
\begin{align}\label{indicesofxys}
	\left[\big[\overline{\boldsymbol{x}}^{(s)}\big]_p,\big[\overline{\boldsymbol{y}}^{(s)}\big]_p\right] = \begin{cases}
		&\!~~~~~~~~~~\mathcal{S}_i\{s\},~~~~~~~~~~~p=\overline{p},\\
		&\!\left[\big[\boldsymbol{x}^{(i-1)}\big]_p,\big[\boldsymbol{y}^{(i-1)}\big]_p\right],~p\neq\overline{p},\\
	\end{cases}
\end{align}
for $p=1,2,\cdots,N_{\rm s}$. Based on $\overline{\boldsymbol{x}}^{(s)}$, $\overline{\boldsymbol{y}}^{(s)}$ and the transformation defined in \eqref{transformation2}, we can obtain the antenna selection matrix $\overline{\overline{\boldsymbol{P}}}^{(s)}$ through 
\begin{align}\label{powerps}
	\overline{\overline{\boldsymbol{P}}}^{(s)} = \mathcal{F}(\overline{\boldsymbol{x}}^{(s)}, \overline{\boldsymbol{y}}^{(s)}).
\end{align}
Given $\overline{\overline{\boldsymbol{P}}}^{(s)}$, (P1) in \eqref{TotalConstraintS1} can be converted to 
\begin{align}\label{designexample}
&\min_{\boldsymbol{F}_{\rm s}}~ \|\mathrm{diag}\left\{\overline{\overline{\boldsymbol{p}}}\right\}\boldsymbol{F}_{\rm s}\|_{\rm F}^2\nonumber\\
&~\mathrm{s.t.}\!~~\eqref{Sub6GSINR},~\eqref{Sub6GBG},
\end{align}
where $\overline{\overline{\boldsymbol{p}}}\triangleq \mathrm{vec}\Big\{\overline{\overline{\boldsymbol{P}}}^{(s)}\Big\}$. Note that \eqref{designexample} is essentially the same as \eqref{subproblemFs}. Therefore, it can be solved following the procedures developed in Section~\ref{OPTFs}. We omit the details and denote the solution as $\overline{\overline{\boldsymbol{F}}}_{\rm s}$. Then, the transmit power of the sub-6G system can be calculated as $\widetilde{P}_{\rm s}^{(s)}$.  We test all the elements in $\mathcal{S}_{i}$ and find the one with the minimum transmit power by
\begin{align}\label{minimumpowerindex}
	s^{*} = \min_{s=1,2,\cdots,|\mathcal{S}_{i}|} \widetilde{P}_{\rm s}^{(s)}.
\end{align}
Then, the row and column indices of the selected antennas in the $i$th iteration can be expressed as 
\begin{align}\label{xiyi}
	\boldsymbol{x}^{(i)} = \overline{\boldsymbol{x}}^{(s^{*})},~\mbox{and}~\boldsymbol{y}^{(i)} = \overline{\boldsymbol{y}}^{(s^{*})}.
\end{align}
Based on $\boldsymbol{x}^{(i)}$, $\boldsymbol{y}^{(i)}$ and the transformation defined in \eqref{transformation2}, we can obtain the antenna selection matrix $\underline{\boldsymbol{P}}^{(i)} = \mathcal{F}(\boldsymbol{x}^{(i)}, \boldsymbol{y}^{(i)})$. The minimum transmit power in the $i$th iteration can be expressed as 
\begin{align}\label{ithminimum1}
	\widehat{P}_{\rm s}^{(i)} = \widetilde{P}_{\rm s}^{(s^{*})}.
\end{align}

We iteratively perform the fast search procedures from \eqref{currentopta} to \eqref{ithminimum1} until one of the following three stop conditions is satisfied.

i) The maximum number of iterations $I_{\rm F}$ is reached;

ii) The minimum transmit power in the $i$th iteration is less than  a predefined power threshold $\breve{P}_{\rm s}$, i.e., $\widehat{P}_{\rm s}^{(i)} \le \breve{P}_{\rm s}$;

iii) The fast search algorithm converges, i.e., $\boldsymbol{x}^{(i)} = \boldsymbol{x}^{(i-p)}$  and $\boldsymbol{y}^{(i)} = \boldsymbol{y}^{(i-p)}$ for $p=1,\cdots, N_{\rm s}-1$.

Denote the number of iterations at this time as $\overline{i}$. Then, the transmit power of the sub-6G system is expressed as  $\overline{P}_{\rm s}=\widehat{P}_{\rm s}^{(\overline{i})}$. Finally, we summarize the developed FS-JBAS algorithm in \textbf{Algorithm~\ref{alg_FSJBAS}}.

We analyze the computational complexity of the FS-JBAS algorithm, which consists of the ABAS algorithm and a fast search procedure. In the ABAS algorithm, $\boldsymbol{F}_{\rm s}$ and $\boldsymbol{p}$ are optimized iteratively for up to $I_{\rm A}$ iterations. The optimization of $\boldsymbol{F}_{\rm s}$ involves solving \eqref{subproblemFs2}, which is an SOCP problem with a computational complexity of $\mathcal{O}(M_{\rm row}^{3.5}M_{\rm col}^{3.5}K_{\rm s}^{3.5})$. Similarly, the optimization of $\boldsymbol{p}$ requires solving \eqref{SubProblemFs3}, which is also an SOCP problem with a complexity of $\mathcal{O}(M_{\rm row}^{3.5}M_{\rm col}^{3.5})$. Therefore, the overall complexity of the ABAS algorithm is $\mathcal{O}(I_{\rm A}M_{\rm row}^{3.5}M_{\rm col}^{3.5}K_{\rm s}^{3.5})$. In the fast search procedure, a maximum of $I_{\rm F}$ iterations are conducted, each involving at most $N_{\rm p}$ tests. Each test solves \eqref{designexample} with a computational complexity of $\mathcal{O}(N_{\rm s}^{3.5}K_{\rm s}^{3.5})$. Consequently, the complexity of the fast search procedure is $\mathcal{O}(I_{\rm F}N_{\rm p}N_{\rm s}^{3.5}K_{\rm s}^{3.5})$. Combining both parts, the total computational complexity of the FS-JBAS algorithm is $\mathcal{O}(K_{\rm s}^{3.5}(I_{\rm F}N_{\rm p}N_{\rm s}^{3.5}+I_{\rm A}M_{\rm row}^{3.5}M_{\rm col}^{3.5}))$.

We also analyze the convergence of the FS-JBAS algorithm, which consists of the ABAS algorithm and a fast search procedure. In the ABAS algorithm, $\boldsymbol{F}_{\rm s}$ and $\boldsymbol{p}$ are iteratively optimized to minimize the objective function. Since the objective value in each iteration does not increase compared to the previous iteration, the convergence of the ABAS algorithm is guaranteed. In the fast search procedure, a total of $I_{\rm F}$ iterations are performed. In each iteration, the solution from the previous iteration serves as the initial value, and the globally optimal solution is obtained through a traversal search method. Consequently, the performance of the current iteration is never worse than that of the previous iteration, ensuring the convergence of the fast search procedure. In summary, the FS-JBAS algorithm converges.

\subsection{Design Example}
Now, we provide an example to facilitate the understanding of the sub-6G beamforming design. As illustrated in Fig.~\ref{fig_DesignExample}, we consider a DBRAA operating at $f_{\rm m} = 30$ GHz and $f_{\rm s} = 5$ GHz, where the numbers of mmWave antennas in  each row and column are both $14$. According to Section \ref{Dual_bandReconAA}, the mmWave antennas can form at most $13\times13$ candidate sub-6G antennas, as shown in Fig.~\ref{fig_DesignExample}a).  We set the minimum antenna spacings as $H_{\rm s} = V_{\rm s} = 6$ to avoid the coupling effects, which implies that the distance between adjacent sub-6G antennas is no less than the half carrier wavelength of sub-6G band. Then, nine candidate sub-6G antennas of (P1-S) in \eqref{SubProblemP1S} are illustrated in Fig.~\ref{fig_DesignExample}b). Suppose $N_{\rm s} = 4$ sub-6G antennas are selected. By solving (P1-S) via \textbf{Algorithm~\ref{alg_JBAS}}, we can select four  antennas from the nine candidate sub-6G antennas, as shown in Fig.~\ref{fig_DesignExample}c). These four antennas are taken as the initial values of the FS-JBAS algorithm. Then, we perform the fast search procedures in Section \ref{DIFSJBAS}, and  four  antennas can be selected from the $13\times13$ candidate sub-6G antennas, as illustrated in Fig.~\ref{fig_DesignExample}d). 

\textbf{Remark 5:} The candidate sub-6G antennas for (P1-S) are the same as those in CAS with the half-wavelength intervals. Using the solution of (P1-S) as the initial values for the FS-JBAS algorithm offers two key advantages. First, by performing the FS-JBAS algorithm to enhance the performance, we can ensure that the performance of the proposed DBRAA will never be worse than that of the conventional sub-6G antenna array with antenna selection, thanks to the convergence of the FS-JBAS algorithm. Second, initializing with the solution of (P1-S) allows the FS-JBAS algorithm to converge more quickly, as will be demonstrated in the simulation~results.

\section{mmWave Beamforming Design}\label{MSD}
In this section, we design the beamforming for mmWave systems by solving (P2) in \eqref{TotalConstraintS2}, where an ADMM-RHB algorithm is developed.

To solve \eqref{TotalConstraintS2}, we first introduce an auxiliary variable $\boldsymbol{F}_{\rm m}$ defined as 
\begin{align}\label{Fm}
	\boldsymbol{F}_{\rm m} = \boldsymbol{F}_{\rm RF}\boldsymbol{F}_{\rm BB}.
\end{align}
Accordingly, the sum-rate in \eqref{Rm} can be rewritten as 
\begin{align}\label{overlineRm}
	\overline{R}_{\rm m} =\sum_{k=1}^{K_{\rm m}}\log_2\left(1 + \frac{|\boldsymbol{h}_{{\rm m},k}^{\rm H}\boldsymbol{f}_{{\rm m},k}|^2}{\sum_{i=1,i\neq k}^{K_{\rm m}}|\boldsymbol{h}_{{\rm m},k}^{\rm H}\boldsymbol{f}_{{\rm m},i}|^2+ \sigma_{\rm m}^2  }\right),
\end{align}
where $\boldsymbol{f}_{{\rm m},k}\triangleq [\boldsymbol{F}_{\rm m}]_{:,k}$. The constraint of the beamforming gain requirements in \eqref{MBGC} can be rewritten as 
\begin{align}\label{MBGC2}
\big\|\boldsymbol{F}_{\rm m}^{\rm H}(\boldsymbol{\alpha}(N_{\rm row},\varphi_{{\rm m},t})\otimes \boldsymbol{\alpha}(N_{\rm col},\vartheta_{{\rm m},t}))\big\|_2\ge \varUpsilon_{{\rm m},t}.
\end{align}
The power constraint in \eqref{mmwavepower} can be rewritten as 
\begin{align}\label{mmwavepower2}
\|\boldsymbol{F}_{\rm m}\|_{\rm F}^2 \le P_{\rm m}.
\end{align}
Thus, \eqref{TotalConstraintS2} is reformulated as 
\begin{subequations}\label{mmWaveDesign1}
\begin{align}
	&\max_{\boldsymbol{F}_{\rm m},\boldsymbol{f}_{\rm P},\boldsymbol{S},\boldsymbol{F}_{\rm BB}}~ \sum_{k=1}^{K_{\rm m}}\overline{R}_{\rm m}\label{maximizationSR}\\
	&~~~~~\mathrm{s.t.}~~~~~~~\eqref{APC},~\eqref{CMC},~\eqref{SC},~\eqref{Fm},~\eqref{MBGC2},~\eqref{mmwavepower2}.
\end{align} 
\end{subequations}
The optimization in \eqref{mmWaveDesign1} is challenging because of the nonconvex log-sum objective in \eqref{maximizationSR}, nonconvex constant modulus constraint in \eqref{CMC}, binary constraint in \eqref{SC}, nonconvex equality constraint in \eqref{Fm}, and nonconvex beamforming gain constraint in \eqref{MBGC2}. In the following, we address these challenging objectives and constraints to solve \eqref{mmWaveDesign1}. 

We first address the nonconvex log-sum objective. Due to the equivalence between the maximization of the sum-rate and the minimization of the WMMSE~\cite{TSP11SQJ}, \eqref{maximizationSR} can be converted~to
\begin{align}
	\min_{u_k,w_k,\boldsymbol{F}_{\rm m},\boldsymbol{f}_{\rm P},\boldsymbol{S},\boldsymbol{F}_{\rm BB}} \sum_{k=1}^{K_{\rm m}}w_ke_k - \log w_k,
\end{align}
where 
\begin{align}\label{ekd}
e_k \triangleq|u_k\boldsymbol{h}_{{\rm m},k}^{\rm H}\boldsymbol{f}_{{\rm m},k} -1|^2 + \sum_{i=1,i\neq k}^{K_{\rm m}}|u_k\boldsymbol{h}_{{\rm m},k}^{\rm H}\boldsymbol{f}_{{\rm m},i}|^2 + \sigma_{\rm m}^2|u_k|^2,
\end{align}
$w_k$ is the weighting coefficients, and $u_k$ is the receiving weights of the $k$th user.

The nonconvex equality constraints in \eqref{APC} and \eqref{Fm} pose great challenges to the optimization of \eqref{mmWaveDesign1}. To streamline the problem, we combine \eqref{APC} and \eqref{Fm} into 
\begin{align}\label{equconstraint}
	\boldsymbol{F}_{\rm m} = {\rm diag}\{\boldsymbol{f}_{\rm P} \}\boldsymbol{S}\boldsymbol{F}_{\rm BB}.
\end{align}
Next, we employ the penalty method to remove the constraint and integrate it into the objective as 
\begin{align}\label{ADMMObjective}
	&~~~~\mathcal{L}(u_k,w_k,\boldsymbol{F}_{\rm m},\boldsymbol{f}_{\rm P},\boldsymbol{S},\boldsymbol{F}_{\rm BB},\boldsymbol{D}) \nonumber\\
	&\triangleq\sum_{k=1}^{K_{\rm m}}w_ke_k - \log w_k + \frac{1}{2\rho}\|\boldsymbol{F}_{\rm m} - {\rm diag}\{\boldsymbol{f}_{\rm P} \}\boldsymbol{S}\boldsymbol{F}_{\rm BB} + \rho\boldsymbol{D}\|_{\rm F}^2,
\end{align}
where $\rho$ is the penalty parameter, and $\boldsymbol{D}$ is the dual variable of the constraint in \eqref{equconstraint}. Thus, \eqref{mmWaveDesign1} can be rewritten as 
\begin{align}\label{mmWaveDesign2}
	&\min_{u_k,w_k,\boldsymbol{F}_{\rm m},\boldsymbol{f}_{\rm P},\boldsymbol{S},\boldsymbol{F}_{\rm BB},\boldsymbol{D}} \mathcal{L}(u_k,w_k,\boldsymbol{F}_{\rm m},\boldsymbol{f}_{\rm P},\boldsymbol{S},\boldsymbol{F}_{\rm BB},\boldsymbol{D}) \nonumber\\
	&~~~~~~~~~~~\mathrm{s.t.}~~~~~~~~~~\eqref{CMC},~\eqref{SC},~\eqref{MBGC2},~\eqref{mmwavepower2}.
\end{align}
Note that \eqref{mmWaveDesign2} is a mixed-integer programming problem, and difficult to solve. Then, we develop an ADMM-RHB algorithm to divide it into several tractable subproblems.

First, to ensure high beamforming gains for communication users and sensing targets, we initialize the ADMM-RHB algorithm by setting $\boldsymbol{F}_{\rm m}$ to $\widetilde{\boldsymbol{F}}_{\rm m}^{(0)}$, where
\begin{align}\label{initializeFm}
\left[\widetilde{\boldsymbol{F}}_{\rm m}^{(0)}\right]_{:,k} \!=\! \kappa\bigg(\!\boldsymbol{h}_{{\rm m},k}\! + \!\sum_{t=1}^{T}\boldsymbol{\alpha}(N_{\rm row},\varphi_{{\rm m},t})\otimes \boldsymbol{\alpha}(N_{\rm col},\vartheta_{{\rm m},t})\bigg).
\end{align}
In \eqref{initializeFm}, $\kappa$ is the scaling factor ensuring the power constraint in \eqref{mmwavepower2} is satisfied. By adopting the manifold optimization-based hybrid beamforming algorithm in \cite{JSTSP16YXH}, we can initialize  $\boldsymbol{F}_{\rm RF}$ and $\boldsymbol{F}_{\rm BB}$ as  $\widetilde{\boldsymbol{F}}_{\rm RF}^{(0)}$ and $\widetilde{\boldsymbol{F}}_{\rm BB}^{(0)}$, respectively. In addition, we initialize $\boldsymbol{D}$ as $\widetilde{\boldsymbol{D}}^{(0)} = \boldsymbol{O}_{N_{\rm p}\times K_{\rm s}}$.

In the $i$th iteration, we sequentially update $u_k$, $w_k$, $\boldsymbol{F}_{\rm m}$, $\boldsymbol{f}_{\rm P}$, $\boldsymbol{S}$, $\boldsymbol{F}_{\rm BB}$, and $\boldsymbol{D}$. For ease of notation, we denote the updated values of $u_k$, $w_k$, $\boldsymbol{F}_{\rm m}$, $\boldsymbol{f}_{\rm P}$, $\boldsymbol{S}$, $\boldsymbol{F}_{\rm BB}$, and $\boldsymbol{D}$ in the $(i-1)$th iteration as $\overline{u}_k$, $\overline{w}_k$, $\boldsymbol{\overline{F}}_{\rm m}$, $\boldsymbol{\overline{f}}_{\rm P}$, $\boldsymbol{\overline{S}}$, $\boldsymbol{\overline{F}}_{\rm BB}$, and $\boldsymbol{\overline{D}}$, respectively.
\subsubsection{Optimization of $u_k$}\label{mmWaveOptuk} When fixing other variables and optimizing $u_k$, \eqref{mmWaveDesign2} can be converted to 
\begin{align}\label{PmmWaveOptuk}
	\min_{u_k} \mathcal{L}\left(u_k,\overline{w}_k,\boldsymbol{\overline{F}}_{\rm m},\boldsymbol{\overline{f}}_{\rm P},\boldsymbol{\overline{S}},\boldsymbol{\overline{F}}_{\rm BB},\boldsymbol{\overline{D}}\right).
\end{align}
Note that \eqref{PmmWaveOptuk} is  an unconstrained optimization problem. Letting $\partial   \mathcal{L}\left(u_k,\overline{w}_k,\boldsymbol{\overline{F}}_{\rm m},\boldsymbol{\overline{f}}_{\rm P},\boldsymbol{\overline{S}},\boldsymbol{\overline{F}}_{\rm BB},\boldsymbol{\overline{D}}\right)/\partial u_k = 0$, we can express the solution of \eqref{PmmWaveOptuk} as 
\begin{align}\label{Solutionuk}
	\widetilde{u}_k^{(i)} = \frac{\boldsymbol{\overline{f}}_{{\rm m},k}^{\rm H}\boldsymbol{h}_{{\rm m},k}}{\sum_{i=1}^{K_{\rm m}}\big|\boldsymbol{\overline{f}}_{{\rm m},i}^{\rm H}\boldsymbol{h}_{{\rm m},k}\big|^2 + \sigma_{\rm m}^2},
\end{align}
where $\overline{\boldsymbol{f}}_{{\rm m},k}\triangleq [\boldsymbol{\overline{F}}_{\rm m}]_{:,k}$.
\subsubsection{Optimization of $w_k$}\label{mmWaveOptwk}
When fixing other variables and optimizing $w_k$, \eqref{mmWaveDesign2} can be converted to 
\begin{align}\label{PmmWaveOptwk}
		\min_{w_k} \mathcal{L}\left(\overline{u}_k,w_k,\boldsymbol{\overline{F}}_{\rm m},\boldsymbol{\overline{f}}_{\rm P},\boldsymbol{\overline{S}},\boldsymbol{\overline{F}}_{\rm BB},\boldsymbol{\overline{D}}\right).
\end{align}
Note that \eqref{PmmWaveOptwk} is also an unconstrained optimization problem. Letting $\partial  \mathcal{L}\left(\overline{u}_k,w_k,\boldsymbol{\overline{F}}_{\rm m},\boldsymbol{\overline{f}}_{\rm P},\boldsymbol{\overline{S}},\boldsymbol{\overline{F}}_{\rm BB},\boldsymbol{\overline{D}}\right)/\partial w_k = 0$, we can express the solution of \eqref{PmmWaveOptwk} as 
\begin{align}\label{wki}
	\widetilde{w}_k^{(i)} = 1/\overline{e}_k,
\end{align}
where $\overline{e}_k$ is obtained by substituting $\widetilde{u}_k^{(i)}$ and $\overline{\boldsymbol{f}}_{{\rm m},k}$ into \eqref{ekd}.
\subsubsection{Optimization of $\boldsymbol{F}_{\rm m}$}\label{mmWaveOptFm} When fixing other variables and optimizing $\boldsymbol{F}_{\rm m}$, \eqref{mmWaveDesign2} can be converted to
\begin{align}\label{PmmWaveOptFm}
    &\min_{\boldsymbol{F}_{\rm m}} \sum_{k=1}^{K_{\rm m}} \widetilde{w}_k^{(i)}\!\left(\! |\widetilde{u}_k^{(i)}\boldsymbol{h}_{{\rm m},k}^{\rm H}\boldsymbol{f}_{{\rm m},k}\! -\! 1|^2\! +\! \sum_{i=1,i\neq k}^{K_{\rm m}}|\widetilde{u}_k^{(i)}\boldsymbol{h}_{{\rm m},k}^{\rm H}\boldsymbol{f}_{{\rm m},i}|^2\!\right) \nonumber\\
    &~~~~~+\frac{1}{2\rho}\|\boldsymbol{F}_{\rm m} - {\rm diag}\{\boldsymbol{\overline{f}}_{\rm P} \}\boldsymbol{\overline{S}}\boldsymbol{\overline{F}}_{\rm BB} + \rho\boldsymbol{\overline{D}}\|_{\rm F}^2 \nonumber\\
    	&~\mathrm{s.t.}~\eqref{MBGC2},~\eqref{mmwavepower2}.
\end{align}
The nonconvex constraint in \eqref{MBGC2} makes \eqref{PmmWaveOptFm}  challenging to solve. Therefore, we develop a successive convex approximation method to solve \eqref{PmmWaveOptFm}.
Similar to \eqref{relax_BG}, the nonconvex beamforming constraint \eqref{MBGC2} can be relaxed to
\begin{align}\label{MBGC3}
	2\mathcal{R}\big\{\boldsymbol{\overline{r}}^{\rm H}\boldsymbol{\overline{B}}\boldsymbol{r}\big\} - \boldsymbol{\overline{r}}^{\rm H}\boldsymbol{\overline{B}}\boldsymbol{\overline{r}}\ge \varUpsilon_{{\rm m},t},
\end{align}
where $\boldsymbol{r}\triangleq \mathrm{vec}\{\boldsymbol{F}_{\rm m}\}$ and $\boldsymbol{\overline{r}}$ denotes the solution of $\boldsymbol{r}$ in the previous iteration, $\boldsymbol{\overline{B}}\triangleq \boldsymbol{I}_{ K_{\rm m}}\otimes\boldsymbol{B}$, and $\boldsymbol{B} = \big(\boldsymbol{\alpha}(N_{\rm row},\varphi_{{\rm m},t})\otimes \boldsymbol{\alpha}(N_{\rm col},\vartheta_{{\rm m},t})\big) \big(\boldsymbol{\alpha}(N_{\rm row},\varphi_{{\rm m},t})\otimes \boldsymbol{\alpha}(N_{\rm col},\vartheta_{{\rm m},t})\big)^{\rm H}$. Then, the approximated convex problem of \eqref{PmmWaveOptFm} can be expressed~as 
\begin{align}\label{PmmWaveOptFm2}
	&\min_{\boldsymbol{F}_{\rm m}} \sum_{k=1}^{K_{\rm m}} \widetilde{w}_k^{(i)}\!\left(\! |\widetilde{u}_k^{(i)}\boldsymbol{h}_{{\rm m},k}^{\rm H}\boldsymbol{f}_{{\rm m},k}\! -\! 1|^2\! +\! \sum_{i=1,i\neq k}^{K_{\rm m}}|\widetilde{u}_k^{(i)}\boldsymbol{h}_{{\rm m},k}^{\rm H}\boldsymbol{f}_{{\rm m},i}|^2\!\right) \nonumber\\
	&~~~~~+\frac{1}{2\rho}\|\boldsymbol{F}_{\rm m} - {\rm diag}\{\boldsymbol{\overline{f}}_{\rm P} \}\boldsymbol{\overline{S}}\boldsymbol{\overline{F}}_{\rm BB} + \rho\boldsymbol{\overline{D}}\|_{\rm F}^2 \nonumber\\
	&~\mathrm{s.t.}~\eqref{mmwavepower2},~\eqref{MBGC3}.
\end{align}
By iteratively solving \eqref{PmmWaveOptFm2}, we can obtain at least a local optimum of \eqref{PmmWaveOptFm}.    We omit the details and denote the solution of \eqref{PmmWaveOptFm} as $\widetilde{\boldsymbol{F}}_{\rm m}^{(i)}$.

\begin{algorithm}[!t]
	\caption{Alternating Direction Method of Multipliers-based Reconfigurable Hybrid Beamforming Algorithm}
	\label{alg_RHB}
	\begin{algorithmic}[1]
		\STATE \textbf{Input:} $N_{\rm row}$, $N_{\rm col}$, $N_{\rm RF}$, $K_{\rm m}$, $T_{\rm m}$,  $\boldsymbol{H}_{\rm m}$, $\varphi_{{\rm m},t}$, and $\vartheta_{{\rm m},t}$.
		\STATE \textbf{Initialization:} Initialize  $\boldsymbol{F}_{\rm m}$ as $\widetilde{\boldsymbol{F}}_{\rm m}^{(0)}$ via \eqref{initializeFm}.
		\STATE ~~~~~~~~~~~~~~~~~~Obtain  $\widetilde{\boldsymbol{F}}_{\rm RF}^{(0)}$ and $\widetilde{\boldsymbol{F}}_{\rm BB}^{(0)}$ based on \cite{JSTSP16YXH}.
		\STATE ~~~~~~~~~~~~~~~~~~$\widetilde{\boldsymbol{D}}^{(0)}\leftarrow\boldsymbol{O}_{N_{\rm p}\times K_{\rm s}}$,~$i\leftarrow1$.
		\WHILE{stop conditions are not satisfied}
		\STATE Obtain $\widetilde{u}_k^{(i)}$ via \eqref{Solutionuk}.
		\STATE Obtain $\widetilde{w}_k^{(i)}$ via \eqref{PmmWaveOptwk}.
		\STATE Obtain $\widetilde{\boldsymbol{F}}_{\rm m}^{(i)}$ via \eqref{PmmWaveOptFm}.
		\STATE Obtain $\widetilde{\boldsymbol{f}}_{\rm P}^{(i)}$ and $\boldsymbol{\widetilde{S}}^{(i)}$ via \eqref{PmmWaveOptFs}.
		\STATE Obtain $\boldsymbol{\widetilde{F}}_{\rm BB}^{(i)}$ via \eqref{FBBsolution}.
		\STATE Obtain  $\boldsymbol{\widetilde{D}}^{(i)}$ via \eqref{updateD}.
		\STATE $i\leftarrow i+1$.
		\ENDWHILE
		\STATE Obtain the current iteration number $\overline{i}$.
		\STATE Obtain $\boldsymbol{\widehat{f}}_{\rm P}$,~$\boldsymbol{\widehat{S}}$, \mbox{and}~$\boldsymbol{\widehat{F}}_{\rm BB}$ via \eqref{mmFinalSolution}.  
		\STATE \textbf{Output:} $\boldsymbol{\widehat{f}}_{\rm P}$,~$\boldsymbol{\widehat{S}}$, \mbox{and}~$\boldsymbol{\widehat{F}}_{\rm BB}$.
	\end{algorithmic}
\end{algorithm}

\subsubsection{Optimization of $\boldsymbol{f}_{\rm P}$ and $\boldsymbol{S}$}\label{mmWaveOptfPS} When fixing other variables, the optimization of $\boldsymbol{f}_{\rm P}$ and $\boldsymbol{S}$ can be expressed as
\begin{align}\label{PmmWaveOptfPfBB}
	&\min_{\boldsymbol{f}_{\rm P},\boldsymbol{S}} \|\boldsymbol{F}_{\rm k} - {\rm diag}\{\boldsymbol{f}_{\rm P} \}\boldsymbol{S}\boldsymbol{\overline{F}}_{\rm BB} \|_{\rm F}^2 \nonumber\\
	&~\mathrm{s.t.}~\eqref{CMC},~\eqref{SC},
\end{align}
where $\boldsymbol{F}_{\rm k}\triangleq \boldsymbol{F}_{\rm m} + \rho\boldsymbol{\overline{D}}$. Note that each row of $\boldsymbol{f}_{\rm P}$ and $\boldsymbol{S}$ is independent of each other. Therefore, we can divide \eqref{PmmWaveOptfPfBB} into $N_{\rm m}$ subproblems, where the $m$th subproblem is expressed as 
\begin{align}\label{PmmWaveOptfPfBB_sub}
	&\min_{f_{{\rm P},m},\boldsymbol{s}_m } \|\boldsymbol{f}_{{\rm k},m} - f_{{\rm P},m}\boldsymbol{\overline{F}}_{\rm BB}^{\rm T}\boldsymbol{s}_m \|_2^2 \nonumber\\
	&~~~\mathrm{s.t.}~~ ~~|f_{{\rm P},m}| =1~\mbox{and}~[\boldsymbol{s}_m]_u \in\{0,1\}.
\end{align}
where $\boldsymbol{f}_{{\rm k},m}\triangleq [\boldsymbol{F}_{\rm k}]_{m,:}^{\rm T}$, $f_{{\rm P},m}\triangleq[\boldsymbol{f}_{\rm P}]_m$,  and $\boldsymbol{s}_m \triangleq [\boldsymbol{S}]_{m,:}^{\rm T}$.  Note that $\boldsymbol{s}_m$ has $N_{\rm RF}$ entries and each entry has only two possible values. Therefore, $\boldsymbol{s}_m$ has totally $2^{N_{\rm RF}}$ possible values. In practice, the number of RF chains is usually small, e.g., four or eight. As a result, the set of possible values for $\boldsymbol{s}_m$ is relatively limited. To this end, we exhaustively test every possible value of $\boldsymbol{s}_m$. We denote the $b$th possible value of $\boldsymbol{s}_m$ as $\boldsymbol{\overline{s}}_m^{(b)}$, for $b=1,2,\cdots,2^{N_{\rm RF}}$.  When testing $\boldsymbol{\overline{s}}_m^{(b)}$, \eqref{PmmWaveOptfPfBB_sub} can be converted as 
\begin{align}\label{PmmWaveOptfPfBB_sub2}
	&\min_{f_{{\rm P},m}} \left\|\boldsymbol{f}_{{\rm k},m} - f_{{\rm P},m}\boldsymbol{\overline{F}}_{\rm BB}^{\rm T}\boldsymbol{\overline{s}}_m^{(b)} \right\|_2^2 \nonumber\\
	&~\mathrm{s.t.}~ |f_{{\rm P},m}| =1.
\end{align}
Note that the objective of \eqref{PmmWaveOptfPfBB_sub2} can be expressed as
\begin{align}\label{PmmWaveOptfPfBB_sub3}
&~~~\|\boldsymbol{f}_{{\rm k},m} - f_{{\rm P},m}\boldsymbol{\overline{F}}_{\rm BB}^{\rm T}\boldsymbol{\overline{s}}_m^{(b)} \|_2^2 \nonumber\\
&=\!\left(\boldsymbol{f}_{{\rm k},m} - f_{{\rm P},m}\boldsymbol{\overline{F}}_{\rm BB}^{\rm T}\boldsymbol{\overline{s}}_m^{(b)}\right)^{\rm H}\left(\boldsymbol{f}_{{\rm k},m} - f_{{\rm P},m}\boldsymbol{\overline{F}}_{\rm BB}^{\rm T}\boldsymbol{\overline{s}}_m^{(b)}\right)\nonumber\\
&=\! \|\boldsymbol{f}_{{\rm k},m}\|_2^2 +\|\boldsymbol{\overline{F}}_{\rm BB}^{\rm T}\boldsymbol{\overline{s}}_m^{(b)}\|_2^2 -2\mathcal{R}\left\{f_{{\rm P},m}\boldsymbol{f}_{{\rm k},m}^{\rm H} \boldsymbol{\overline{F}}_{\rm BB}^{\rm T}\boldsymbol{\overline{s}}_m^{(b)}\right\}.
\end{align}
Since $\boldsymbol{f}_{{\rm k},m}$ and $\overline{\boldsymbol{F}}_{\rm BB}^{\rm T}\boldsymbol{\overline{s}}_m^{(b)}$ are irrelevant to $f_{{\rm P},m}$, the optimal solution of \eqref{PmmWaveOptfPfBB_sub2} can be expressed as 
\begin{align}
	\overline{f}_{{\rm P},m}^{(b)} = {\rm exp}\left\{-j\mathcal{A}\left(\boldsymbol{f}_{{\rm k},m}^{\rm H} \boldsymbol{\overline{F}}_{\rm BB}^{\rm T}\boldsymbol{\overline{s}}_m^{(b)}\right)\right\},
\end{align}
where $\mathcal{A}(\cdot)$ denotes the phase of a complex number. Then, the objective of \eqref{PmmWaveOptfPfBB_sub2} can be calculated as 
\begin{align}
C_{m}^{(b)}  = \left\|\boldsymbol{f}_{{\rm k},m} - \overline{f}_{{\rm P},m}^{(b)}\boldsymbol{\overline{F}}_{\rm BB}^{\rm T}\boldsymbol{\overline{s}}_m^{(b)} \right\|_2^2.
\end{align}
We select the value of $\boldsymbol{s}_{m}$ that minimizes the objective by
\begin{align}
	b^{*} = \min_{b=1,2,\cdots,2^{N_{\rm RF}}}~C_{m}^{(b)}.
\end{align}
Then, the optimal solution of \eqref{PmmWaveOptfPfBB_sub} can be expressed as
\begin{align}
	\widetilde{f}_{{\rm P},m} = \overline{f}_{{\rm P},m}^{(b^{*})},~\mbox{and}~\widetilde{\boldsymbol{s}}_m = \boldsymbol{\overline{s}}_m^{(b^{*})}.
\end{align}
By combining the solutions of the $N_{\rm p}$ subproblems in \eqref{PmmWaveOptfPfBB_sub}, the optimal solution of \eqref{PmmWaveOptfPfBB} can be expressed as
\begin{align}\label{PmmWaveOptFs}
	&\widetilde{\boldsymbol{f}}_{\rm P}^{(i)} = \left[\widetilde{f}_{{\rm P},1},\widetilde{f}_{{\rm P},2},\cdots,\widetilde{f}_{{\rm P},N_{\rm p}}\right]^{\rm T},\nonumber\\
	&\boldsymbol{\widetilde{S}}^{(i)} = \left[\widetilde{\boldsymbol{s}}_1,\widetilde{\boldsymbol{s}}_2,\cdots,\widetilde{\boldsymbol{s}}_{ N_{\rm p}}\right]^{\rm T}.
\end{align}
\subsubsection{Optimization of $\boldsymbol{F}_{\rm BB}$}\label{mmWaveOptFBB}When fixing other variables and optimizing $\boldsymbol{F}_{\rm BB}$, \eqref{mmWaveDesign2} can be converted to
\begin{align}\label{FBBopt}
\min_{\boldsymbol{F}_{\rm BB}} \left\|\boldsymbol{F}_{\rm k} - {\rm diag}\left\{\widetilde{\boldsymbol{f}}_{\rm P}^{(i)}\right\}\boldsymbol{\widetilde{S}}^{(i)}\boldsymbol{F}_{\rm BB} \right\|_{\rm F}^2.
\end{align}
Note that \eqref{FBBopt} is an unconstrained least-squares problem and its optimal solution  can be expressed as 
\begin{align}\label{FBBsolution}
\boldsymbol{\widetilde{F}}_{\rm BB}^{(i)} =  \left(\widetilde{\boldsymbol{F}}_{\rm RF}^{(i){\rm H}}\widetilde{\boldsymbol{F}}_{\rm RF}^{(i)}\right)^{-1}\widetilde{\boldsymbol{F}}_{\rm RF}^{(i){\rm H}}\boldsymbol{F}_{\rm k},
\end{align}
where $\widetilde{\boldsymbol{F}}_{\rm RF}^{(i)} = {\rm diag}\big\{\widetilde{\boldsymbol{f}}_{\rm P}^{(i)}\big\}\boldsymbol{\widetilde{S}}^{(i)}$.
\subsubsection{Updates of $\boldsymbol{D}$}
According to the convention of the ADMM, we update the dual variables, $\boldsymbol{D}$, via
\begin{align}\label{updateD}
	 \boldsymbol{\widetilde{D}}^{(i)} =   \boldsymbol{\overline{D}} + \frac{1}{\rho}\big(\widetilde{\boldsymbol{F}}_{\rm m}^{(i)} - \widetilde{\boldsymbol{F}}_{\rm RF}^{(i)}\widetilde{\boldsymbol{F}}_{\rm BB}^{(i)}\big).
\end{align}

We iteratively perform the procedures from \eqref{PmmWaveOptwk} to \eqref{updateD} until one of the following two stop conditions is satisfied.

i) The predefined maximum number of iterations $I_{\rm M}$ is reached.

ii) The disparity  between the current beamformer and the fomer one is less than a predefined threshold $\epsilon$, i.e., $\left\|\widetilde{\boldsymbol{F}}_{\rm RF}^{(i)}\widetilde{\boldsymbol{F}}_{\rm BB}^{(i)} - \widetilde{\boldsymbol{F}}_{\rm RF}^{(i-1)}\widetilde{\boldsymbol{F}}_{\rm BB}^{(i-1)}\right\|_{\rm F}^2\le\epsilon$.

Denote the number of iterations at this time as $\overline{i}$. Then, the final solutions of (P2) are expressed as
\begin{align}\label{mmFinalSolution}
\boldsymbol{\widehat{f}}_{\rm P} = \widetilde{\boldsymbol{f}}_{\rm P}^{(\overline{i})},~\boldsymbol{\widehat{S}} = \boldsymbol{\widetilde{S}}^{(\overline{i})},\mbox{and}~\boldsymbol{\widehat{F}}_{\rm BB} = \boldsymbol{\widetilde{F}}_{\rm BB}^{(\overline{i})}.  
\end{align}

\begin{figure*}[htbp]
	\centering
	\includegraphics[width=180mm]{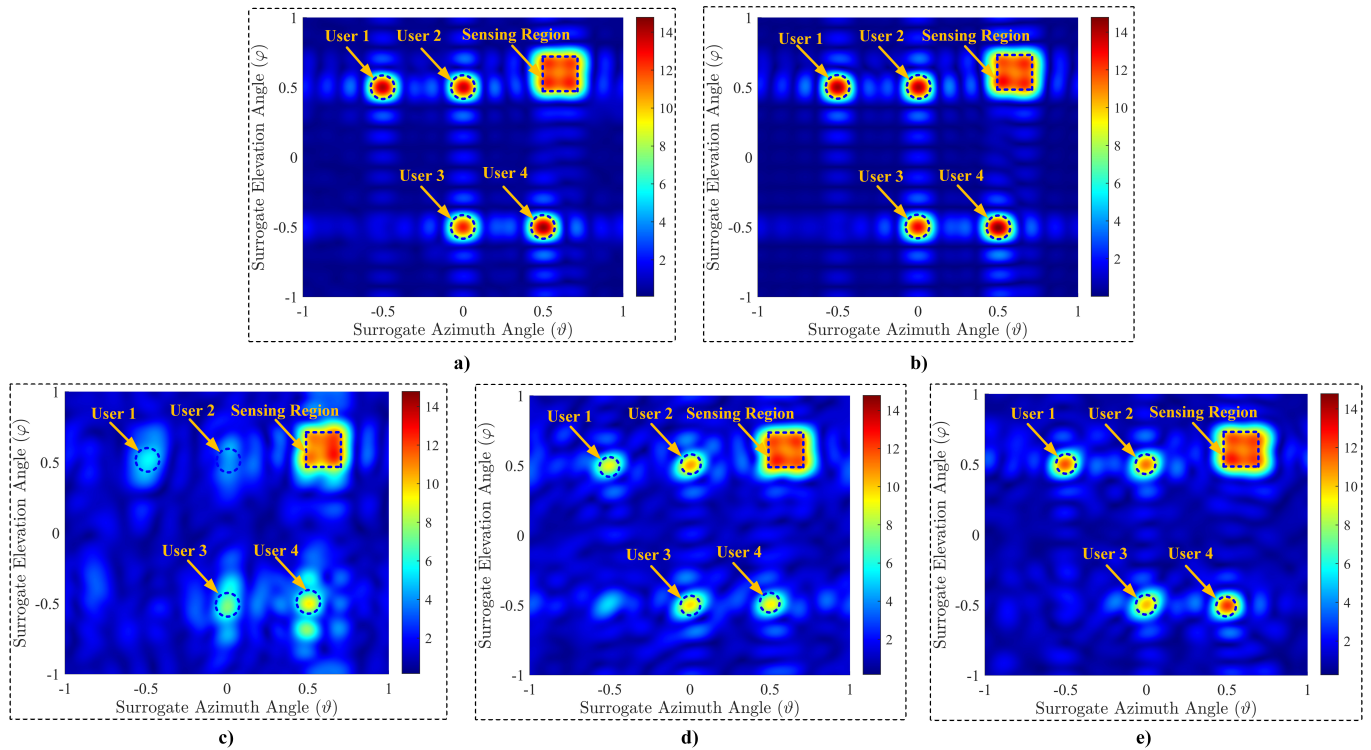}
	\caption{Comparisons of the beamforming designs with different structures in terms of the beamforming gains: a) Fully digital structure; b) Fully-connected hybrid beamforming structure; c)  Partially-connected hybrid beamforming structure; d) Dynamic hybrid beamforming structure; e) Reconfigurable hybrid beamforming structure.}
	\label{Fig_BeamDesign}
\end{figure*}

Finally, we summarize the developed ADMM-RHB algorithm in \textbf{Algorithm~\ref{alg_RHB}}.

Now, we analyze the computational complexity of the ADMM-RHB algorithm. The ADMM-RHB algorithm involves at most $I_{\rm M}$ iterations. In each iteration, the algorithm sequentially optimizes $u_k$, $w_k$, $\boldsymbol{F}_{\rm m}$, $\boldsymbol{f}_{\rm P}$, $\boldsymbol{S}$, $\boldsymbol{F}_{\rm BB}$, and $\boldsymbol{D}$. The optimization of $u_k$ in \eqref{Solutionuk} has a computational complexity of $\mathcal{O}(K_{\rm m}N_{\rm m})$, while the optimization of $w_k$ in \eqref{wki} has a computational complexity of $\mathcal{O}(K_{\rm m})$. The optimization of $\boldsymbol{F}_{\rm m}$ is performed by iteratively solving \eqref{PmmWaveOptFm2}, which is a quadratic programming problem with a computational complexity of $\mathcal{O}(N_{\rm m}^{3.5}K_{\rm m}^{3.5})$. Let $I_{\rm S}$ denote the maximum number of iterations required to solve \eqref{PmmWaveOptFm2}. Thus, the computational complexity of optimizing $\boldsymbol{F}_{\rm m}$ is $\mathcal{O}(I_{\rm S}N_{\rm m}^{3.5}K_{\rm m}^{3.5})$. The optimization of $\boldsymbol{f}_{\rm P}$ is performed entry-wise, with $2^{N_{\rm RF}}$ tests of $\boldsymbol{s}_m$ conducted for each entry. Therefore, the computational complexity of optimizing $\boldsymbol{f}_{\rm P}$ and $\boldsymbol{S}$ is $\mathcal{O}(N_{\rm m}2^{N_{\rm RF}})$. The optimization of $\boldsymbol{F}_{\rm BB}$ in \eqref{FBBsolution} has a computational complexity of $\mathcal{O}(N_{\rm m}N_{\rm RF} + N_{\rm RF}^3)$. The update of $\boldsymbol{D}$ is performed using a closed-form expression, and its computational complexity is negligible. In summary, the computational complexity of the ADMM-RHB algorithm is $\mathcal{O}(I_{\rm M}(I_{\rm S}N_{\rm m}^{3.5}K_{\rm m}^{3.5} + N_{\rm m}2^{N_{\rm RF}}))$.

We also analyze the convergence of the ADMM-RHB algorithm. The algorithm iteratively optimizes $u_k$, $w_k$, $\boldsymbol{F}_{\rm m}$, $\boldsymbol{f}_{\rm P}$, $\boldsymbol{S}$, $\boldsymbol{F}_{\rm BB}$, and $\boldsymbol{D}$ to minimize the objective function in \eqref{ADMMObjective}. In each optimization step, either a closed-form solution is obtained or the variables are iteratively updated based on prior values. As a result, the objective function is guaranteed to be monotonically non-increasing throughout the optimization process. Furthermore, through the iterative dual variable updates in ADMM, the enforcement of constraints is progressively strengthened. Consequently, the hybrid beamformer $\boldsymbol{F}_{\rm RF}\boldsymbol{F}_{\rm BB}$ asymptotically approaches the fully-digital beamformer $\boldsymbol{F}_{\rm m}$. In summary, the ADMM-RHB algorithm converges as the objective function decreases monotonically and the constraints in \eqref{equconstraint} are increasingly enforced.

\section{Simulation Results}\label{SR}
Now we evaluate the performance of the dual-band ISAC systems with the DBRAA. The sub-6G system operates at $f_{\rm s} = 5$ GHz and the mmWave system operates at $f_{\rm m} = 30$~GHz, which implies that the carrier wavelength of the former is six times that of the latter. Therefore, we set the minimum antenna spacings as $H_{\rm s} = V_{\rm s} = 6$. We set  $L_{{\rm m},k} = 3$  and $L_{{\rm s},k} = 5$  for the mmWave and sub-6G channels, respectively.   The noise power of both systems is normalized as  $\sigma^2_{\rm m} = \sigma^2_{\rm s} = 1$.

\subsection{Evaluating the Performance of mmWave Systems}\label{EMS}
First, we evaluate the performance of the mmWave systems with the RHB structure.

In Fig.~\ref{Fig_BeamDesign}, we compare the beamforming performance of different beamforming structures in terms of the beamforming gains. The mmWave antenna array is set with $N_{\rm row} = N_{\rm col} = 14$, resulting in $N_{\rm m} = 196$ antennas in total. Both the number of mmWave users and the number of RF chains are set to four. The sensing targets are located in a region defined by surrogate azimuth and elevation angles ranging from 0.5 to 0.7. The minimum sensing beamforming gain is set to $\varUpsilon_{{\rm m},t} = 11$. The mmWave transmit power is set to $7$. From the figure, the FD structure demonstrates the best performance for both sensing and communications due to its maximum design flexibility. The fully-connected hybrid beamforming (FCHB) structure closely approaches the performance of the FD structure but shows slight distortion in the sensing region. The partially-connected hybrid beamforming (PCHB) structure exhibits the worst performance among all methods due to its limited design flexibility. The DHB structure significantly outperforms the PCHB structure due to the introduction of a switch network. Nonetheless, it performs worse than the FCHB structure because it uses far fewer phase shifters. The RHB structure, which offers higher design flexibility in the switch network than the DHB structure, achieves higher beamforming gains for communications and a more regular sensing region.

\begin{figure}[!t]
	\centering
	\includegraphics[width=70mm]{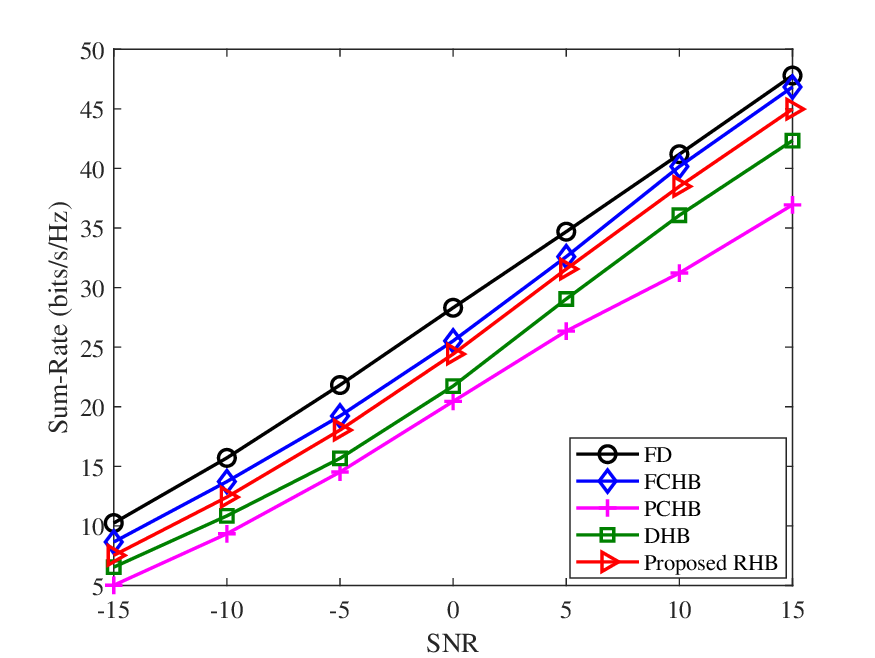}
	\caption{Comparisons of different mmWave beamforming structures in terms of the sum-rate for different SNRs. }
	\label{Fig_SumRate}
\end{figure}

In Fig.~\ref{Fig_SumRate}, we compare different beamforming structures in terms of the sum-rate for different SNRs. The antenna array, the number of mmWave users, and the number of RF chains are configured the same as those in Fig.~\ref{Fig_BeamDesign}. The results show that the FD exhibits the best performance, followed by the FCHB, the RHB, and the DHB, with the PCHB showing the worst  performance. Notably, the proposed RHB structure outperforms the DHB structure and closely approaches the performance of the FCHB structure, confirming the effectiveness of the proposed method.

\begin{figure}[!t]
	\centering
	\includegraphics[width=70mm]{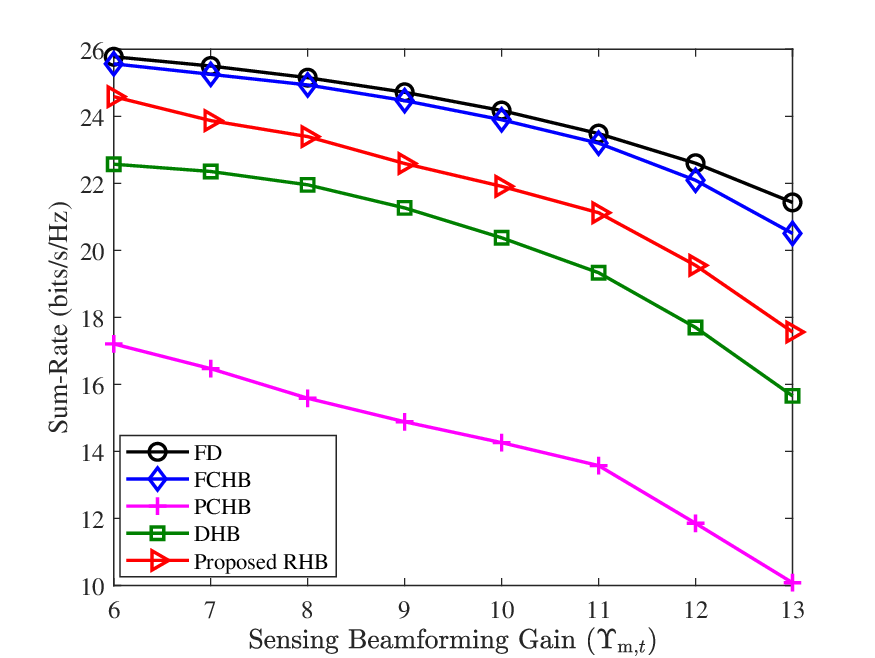}
	\caption{Trade-off between the communication sum-rate and  sensing beamforming gain for different mmWave beamforming structures. }
	\label{Fig_TradeOffSC}
\end{figure}

In Fig.~\ref{Fig_TradeOffSC}, we illustrate the trade-off between the communication sum-rate and the sensing beamforming gain for different mmWave beamforming structures, where the antenna array, the number of  mmWave users, the number of RF chains, and the transmit power are set the  same as those in  Fig.~\ref{Fig_BeamDesign}. From the figure, the communication sum-rate decreases with the increase of the sensing beamforming gain. This is because the total power of sensing and  communications is fixed, and the increase in  power consumed by sensing will lead to  the decrease in the communication performance. In addition, the proposed RHB  performs better than the DHB with similar hardware complexity, which verifies again the  effectiveness of the proposed method.


\begin{figure}[t]
	\centering
	\includegraphics[width=70mm]{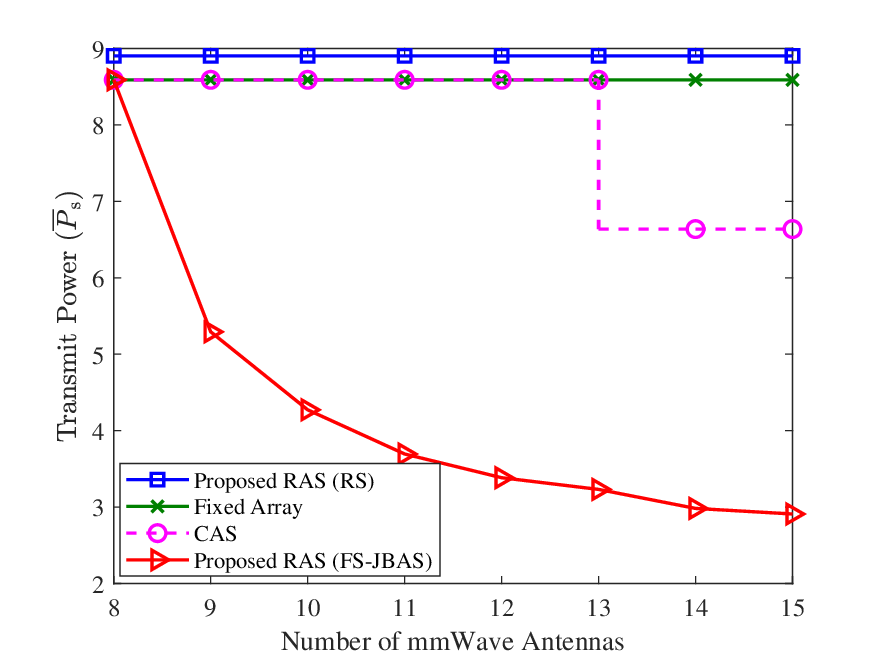}
	\caption{Comparisons of different sub-6G systems for different numbers of mmWave antennas. }
	\label{Fig_ChangeWithNumberAntennas}
\end{figure}

\begin{figure}[t]
	\centering
	\includegraphics[width=70mm]{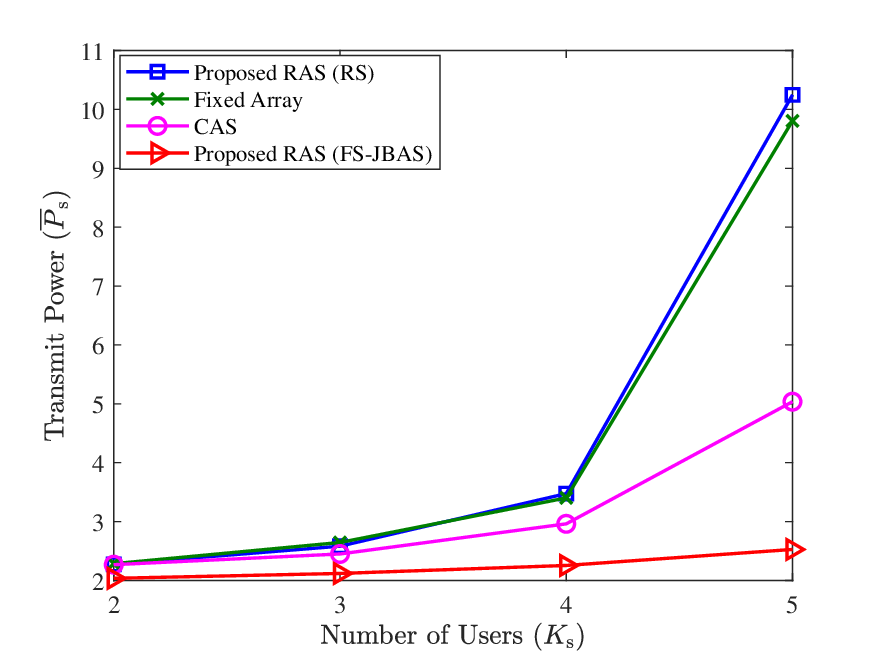}
	\caption{Comparisons of different sub-6G systems for different numbers of users. }
	\label{Fig_ChangeWithNumberofUsers}
\end{figure}

\begin{figure}[t]
	\centering
	\includegraphics[width=70mm]{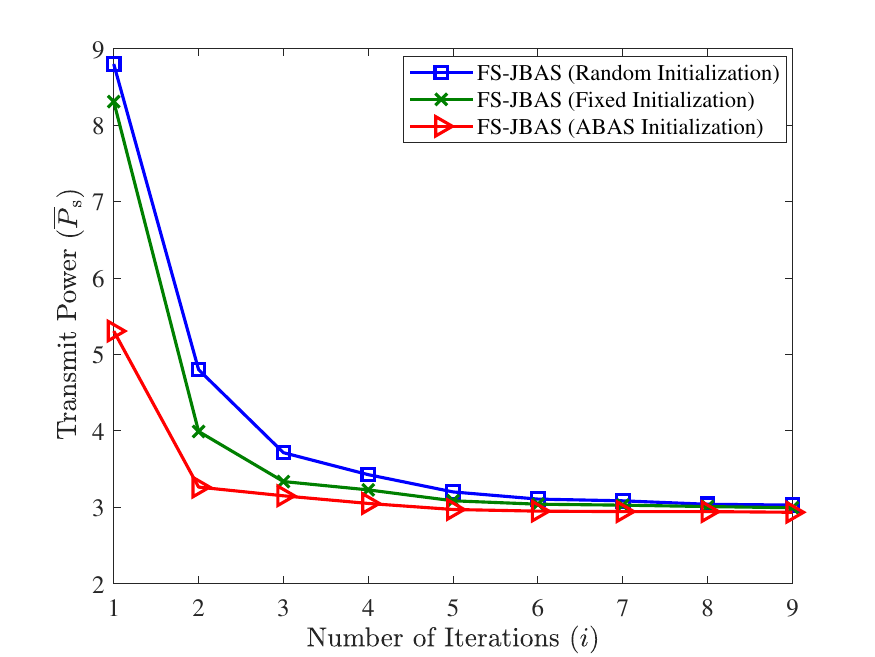}
	\caption{Comparisons of different initialization methods for the FS-JBAS algorithm. }
	\label{Fig_ChangeWithNumberofIteration}
\end{figure}


\subsection{Evaluating the Performance of Sub-6 GHz Systems}\label{ESS}
In the following, we evaluate the performance of the sub-6G systems with the RAS structure. 

In Fig.~\ref{Fig_ChangeWithNumberAntennas}, we compare the sub-6G systems with different array structures, where the number of sub-6G antennas is $N_{\rm s} = 4$. The horizontal axis represents the number of antennas in each row and column of the mmWave antenna array used for forming sub-6G antennas. The CAS, which has the same aperture as the mmWave antenna array with half-wavelength spacing between adjacent candidate antennas, is used as a benchmark.  Another benchmark is the fixed array, which has fixed $N_{\rm s}$ antennas with half-wavelength spacings. When $N_{\rm row}$ and $N_{\rm col}$ are less than 13, the array panel can only hold four candidate sub-6G antennas, resulting in CAS performing the same as the fixed array. The minimum sensing beamforming gain, $\Upsilon_{{\rm s},t}$, and the SINR threshold, $\Gamma_{k}$, are both set to $10$.  From the figure, as  $N_{\rm row}$ and $N_{\rm row}$ increase, the consumed power of the proposed RAS decreases due to the increased design flexibility. Additionally, the RAS significantly outperforms the CAS and fixed array, which verifies the effectiveness of the proposed RAS structure. Furthermore, the RAS with the FS-JBAS algorithm significantly outperforms the RAS with random selection (RS), which demonstrates the effectiveness of the proposed FS-JBAS algorithm.

In Fig.~\ref{Fig_ChangeWithNumberofUsers}, we compare the sub-6G systems with different array structures in terms of different numbers of users, where the number of sub-6G antennas is $N_{\rm s} = 5$. The  mmWave antenna array is configured with $N_{\rm row} = N_{\rm col} = 14$.  The minimum sensing beamforming gain, $\Upsilon_{{\rm s},t}$, and the SINR threshold, $\Gamma_{k}$, are both set to $10$. The figure shows that when the number of users is fewer than four, all methods require low transmit power. However, when the number of users reaches five, the transmit powers for RAS (RS), fixed array, and CAS increase significantly, whereas the transmit power for RAS (FS-JBAS) only increases slightly. This is because the RAS enables more effective multiuser interference mitigation by finely adjusting the antenna positions, making it more advantageous in complex scenarios.

In Fig.~\ref{Fig_ChangeWithNumberofIteration}, we compare different initialization methods for the FS-JBAS algorithm, where  the sub-6G array is set as  $N_{\rm s} = 4$ and the mmWave array is set as $N_{\rm row} = N_{\rm col} = 14$. The number of sub-6G users is set as $K_{\rm s} = 4$. The minimum sensing beamforming gain, $\Upsilon_{{\rm s},t}$, and the SINR threshold, $\Gamma_{k}$, are both set to $10$. The random initialization, which randomly configures sub-6G antennas to initialize the FS-JBAS algorithm, and the fixed initialization, which initializes the FS-JBAS algorithm using a fixed array, are used as benchmarks. The figure demonstrates that ABAS initialization outperforms both random and fixed initialization in terms of convergence speed and performance, verifying the effectiveness of the proposed ABAS initialization method.

\section{Conclusion}\label{CC}
In this paper, a DBRAA,  enabling wireless capabilities in both sub-6G and mmWave bands using a single array, has been proposed. An RAS structure and an RHB  structure have been developed for the sub-6G and mmWave bands, respectively. ISAC in sub-6G and mmWave bands using the DBRAA has been investigated and a dual-band ISAC beamforming design problem has been formulated. An FS-JBAS algorithm and an ADMM-RHB algorithm have been developed for the sub-6G and mmWave beamforming designs, respectively.  Simulation results have shown that the RAS structure outperforms the conventional CAS structure and RHB  structure outperforms the existing DHB structure. In future work, we will continue to investigate the performance of DBRAA for sensing and communications, with a primary focus on developing efficient beamforming algorithms for real-time implementation.

\section{Acknowledgment}\label{Ac}
The authors would like to thank Mr. Yujing Hong  from Nanyang Technological University for his valuable help in discussing the key antenna techniques related to the DBRAA.

\bibliographystyle{IEEEtran}
\bibliography{IEEEabrv,IEEEexample}

\begin{thebibliography}{10}
\providecommand{\url}[1]{#1}
\csname url@samestyle\endcsname
\providecommand{\newblock}{\relax}
\providecommand{\bibinfo}[2]{#2}
\providecommand{\BIBentrySTDinterwordspacing}{\spaceskip=0pt\relax}
\providecommand{\BIBentryALTinterwordstretchfactor}{4}
\providecommand{\BIBentryALTinterwordspacing}{\spaceskip=\fontdimen2\font plus
\BIBentryALTinterwordstretchfactor\fontdimen3\font minus
  \fontdimen4\font\relax}
\providecommand{\BIBforeignlanguage}[2]{{%
\expandafter\ifx\csname l@#1\endcsname\relax
\typeout{** WARNING: IEEEtran.bst: No hyphenation pattern has been}%
\typeout{** loaded for the language `#1'. Using the pattern for}%
\typeout{** the default language instead.}%
\else
\language=\csname l@#1\endcsname
\fi
#2}}
\providecommand{\BIBdecl}{\relax}
\BIBdecl

\bibitem{GlobeCom25CKJ}
K.~Chen, C.~Qi, and O.~A. Dobre, ``Sub-6 {GHz} and millimeter wave dual-band
  reconfigurable antenna array,'' in \emph{Proc. IEEE Global Commun. Conf.
  (GLOBECOM), to be submitted}, Taipei, Taiwan, Dec. 2025, pp. 1--6.

\bibitem{yang2019}
P.~Yang, Y.~Xiao, M.~Xiao, and S.~Li, ``{6G} wireless communications: {Vision}
  and potential techniques,'' \emph{IEEE Netw.}, vol.~33, no.~4, pp. 70--75,
  July 2019.

\bibitem{OBETrans}
Z.~Wang, J.~Zhang, E.~Bj{\"o}rnson, D.~Niyato, and B.~Ai, ``Optimal bilinear
  equalizer for cell-free massive {MIMO} systems over correlated {R}ician
  channels,'' \emph{IEEE Trans. Signal Process.}, pp. 1--16, to appear, 2025.

\bibitem{IoT2024LSH}
S.~Lu \emph{et~al.}, ``Integrated sensing and communications: {Recent} advances
  and ten open challenges,'' \emph{IEEE Internet Things J.}, vol.~11, no.~11,
  pp. 19\,094--19\,120, June 2024.

\bibitem{Tcom24WZQ}
Z.~Wei \emph{et~al.}, ``Waveform design for {MIMO-OFDM} integrated sensing and
  communication system: {An} information theoretical approach,'' \emph{{IEEE}
  Trans. Commun.}, vol.~72, no.~1, pp. 496--509, Jan. 2024.

\bibitem{Tcom20Liufan}
F.~Liu \emph{et~al.}, ``Joint radar and communication design: {Applications},
  state-of-the-art, and the road ahead,'' \emph{{IEEE} Trans. Commun.},
  vol.~68, no.~6, pp. 3834--3862, Feb. 2020.

\bibitem{JSTSP18Zhengle}
L.~Zheng, M.~Lops, and X.~Wang, ``Adaptive interference removal for
  uncoordinated radar/communication coexistence,'' \emph{{IEEE} J. Sel. Top.
  Signal Process.}, vol.~12, no.~1, pp. 45--60, Feb. 2018.

\bibitem{TWC23CKJ1}
K.~Chen, C.~Qi, O.~A. Dobre, and G.~Y. Li, ``Simultaneous beam training and
  target sensing in {ISAC} systems with {RIS},'' \emph{{IEEE} Trans. Wireless
  Commun.}, vol.~23, no.~4, pp. 2696--2710, Apr. 2024.

\bibitem{IEEENetwork24_DFW}
F.~Dong, F.~Liu, Y.~Cui, S.~Lu, and Y.~Li, ``Sensing as a service in {6G}
  perceptive mobile networks: {Architecture}, advances, and the road ahead,''
  \emph{IEEE Netw.}, vol.~38, no.~2, pp. 87--96, Mar. 2024.

\bibitem{TWC23DFW}
F.~Dong, F.~Liu, Y.~Cui, W.~Wang, K.~Han, and Z.~Wang, ``Sensing as a service
  in {6G} perceptive networks: {A} unified framework for {ISAC} resource
  allocation,'' \emph{{IEEE} Trans. Wireless Commun.}, vol.~22, no.~5, pp.
  3522--3536, May 2023.

\bibitem{CL24CKJ}
K.~Chen and C.~Qi, ``Joint sparse {Bayesian} learning for channel estimation in
  {ISAC},'' \emph{IEEE Commun. Lett.}, vol.~28, no.~8, Aug. 2024.

\bibitem{TWC24WY}
Y.~Wang, Q.~Zhang, J.~Andrew~Zhang, Z.~Wei, Z.~Feng, and J.~Peng,
  ``Interference characterization and mitigation for multi-beam {ISAC} systems
  in vehicular networks,'' \emph{{IEEE} Trans. Wireless Commun.}, vol.~23,
  no.~10, pp. 14\,729--14\,742, Oct. 2024.

\bibitem{JSAC17SM}
M.~Shafi \emph{et~al.}, ``{5G}: {A} tutorial overview of standards, trials,
  challenges, deployment, and practice,'' \emph{{IEEE} J. Sel. Areas Commun.},
  vol.~35, no.~6, June 2017.

\bibitem{CL22Chenhao}
C.~Qi, W.~Ci, J.~Zhang, and X.~You, ``Hybrid beamforming for millimeter wave
  {MIMO} integrated sensing and communications,'' \emph{IEEE Commun. Lett.},
  vol.~26, no.~5, pp. 1136--1140, May 2022.

\bibitem{TVT20CM}
M.~Cheng, J.-B. Wang, J.~Cheng, J.-Y. Wang, and M.~Lin, ``Joint scheduling and
  precoding for {mmWave} and {Sub-6 GHz} dual-mode networks,'' \emph{{IEEE}
  Trans. Veh. Technol.}, vol.~69, no.~11, pp. 13\,098--13\,111, Nov. 2020.

\bibitem{Balanis2015}
C.~A. Balanis, \emph{Antenna Theory: Analysis and Design}, 4th~ed.\hskip 1em
  plus 0.5em minus 0.4em\relax Hoboken, NJ, USA: Wiley, 2016.

\bibitem{TAP23_SL}
L.~Sang \emph{et~al.}, ``A dual-band planar antenna array with high-frequency
  ratio for both {Sub-6} band and {mm-Wave} band,'' \emph{{IEEE} Trans.
  Antennas Propag.}, vol.~71, no.~5, pp. 3856--3867, May 2023.

\bibitem{AWPL24TXY}
X.~Tong, J.~Xu, H.~Zhang, and X.~Yu, ``An integrated microwave and
  millimeter-wave circularly polarized antenna with {ISM} band gain suppression
  performance,'' \emph{{IEEE} Antennas Wireless Propag. Lett.}, vol.~23,
  no.~11, pp. 3649--3653, Nov. 2024.

\bibitem{liu2023joint}
R.~Liu, M.~Li, and Q.~Liu, ``Joint transmit/receive antenna selection and
  beamforming design for {ISAC} systems,'' in \emph{Proc. IEEE Global Commun.
  Conf. (GLOBECOM)}, Kuala Lumpur, Malaysia, 2023, pp. 3118--3123.

\bibitem{TWC24MWY}
W.~Ma, L.~Zhu, and R.~Zhang, ``{MIMO} capacity characterization for movable
  antenna systems,'' \emph{{IEEE} Trans. Wireless Commun.}, vol.~23, no.~4, pp.
  3392--3407, Apr. 2024.

\bibitem{TWC21KKW}
K.-K. Wong, A.~Shojaeifard, K.-F. Tong, and Y.~Zhang, ``Fluid antenna
  systems,'' \emph{{IEEE} Trans. Wireless Commun.}, vol.~20, no.~3, pp.
  1950--1962, Mar. 2021.

\bibitem{WC24_YKK}
K.~Ying \emph{et~al.}, ``Reconfigurable massive {MIMO}: {Harnessing} the power
  of the electromagnetic domain for enhanced information transfer,'' \emph{IEEE
  Wireless Commun.}, vol.~31, no.~3, pp. 125--132, June 2024.

\bibitem{APM13_HR}
R.~L. Haupt and M.~Lanagan, ``Reconfigurable antennas,'' \emph{{IEEE} Antennas
  Propag. Mag.}, vol.~55, no.~1, pp. 49--61, Feb. 2013.

\bibitem{IEEE_15_P}
J.~Costantine, Y.~Tawk, S.~E. Barbin, and C.~G. Christodoulou, ``Reconfigurable
  antennas: {Design} and applications,'' \emph{Proc. IEEE}, vol. 103, no.~3,
  pp. 424--437, Mar. 2015.

\bibitem{zhang2024pixel}
J.~Zhang \emph{et~al.}, ``A novel pixel-based reconfigurable antenna applied in
  fluid antenna systems with high switching speed,'' \emph{{IEEE} Open J.
  Antennas Propag.}, vol.~6, no.~1, pp. 212--228, Feb. 2025.

\bibitem{TAP22_JLW}
L.~Jing, M.~Li, and R.~Murch, ``Compact pattern reconfigurable pixel antenna
  with diagonal pixel connections,'' \emph{{IEEE} Trans. Antennas Propag.},
  vol.~70, no.~10, pp. 8951--8961, Oct. 2022.

\bibitem{TAP17_LP}
P.~Lotfi, S.~Soltani, and R.~D. Murch, ``Printed endfire beam-steerable pixel
  antenna,'' \emph{{IEEE} Trans. Antennas Propag.}, vol.~65, no.~8, pp.
  3913--3923, Aug. 2017.

\bibitem{Rao2024}
J.~Rao \emph{et~al.}, ``A shared-aperture dual-band sub-6 {GHz} and {mmWave}
  reconfigurable intelligent surface with independent operation,'' \emph{IEEE
  Trans. Microw. Theory Tech., early access}, pp. 1--17, 2024.

\bibitem{TAP14HSV}
S.~V. Hum and J.~Perruisseau-Carrier, ``Reconfigurable reflectarrays and array
  lenses for dynamic antenna beam control: {A} review,'' \emph{{IEEE} Trans.
  Antennas Propag.}, vol.~62, no.~1, pp. 183--198, Jan. 2014.

\bibitem{Saifullah2022}
Y.~Saifullah, Y.~He, A.~Boag, G.-M. Yang, and F.~Xu, ``Recent progress in
  reconfigurable and intelligent metasurfaces: {A} comprehensive review of
  tuning mechanisms, hardware designs, and applications,'' \emph{Adv. Sci.},
  vol.~9, no.~30, p. 2203747, Sep. 2022.

\bibitem{TAP24ZLH}
L.~Zhu \emph{et~al.}, ``Dual linearly polarized 2-bit programmable metasurface
  with high cross-polarization discrimination,'' \emph{{IEEE} Trans. Antennas
  Propag.}, vol.~72, no.~2, pp. 1510--1520, Feb. 2024.

\bibitem{Hong2024}
Y.~Hong, Y.~Zhao, C.~Yuen, and X.~Qing, ``A {STAR-RIS} with independent 1-bit
  wave control,'' in \emph{Proc. 2024 IEEE Int. Symp. Antennas Propag.
  INC/USNC-URSI Radio Sci. Meeting}, Firenze, Italy, 2024, pp. 813--814.

\bibitem{APM15VM}
M.~Vogel, G.~Gampala, D.~Ludick, U.~Jakobus, and C.~J. Reddy, ``Characteristic
  mode analysis: {Putting} physics back into simulation,'' \emph{{IEEE}
  Antennas Propag. Mag.}, vol.~57, no.~2, pp. 307--317, Apr. 2015.

\bibitem{TAP17LF}
F.~H. Lin and Z.~N. Chen, ``Low-profile wideband metasurface antennas using
  characteristic mode analysis,'' \emph{{IEEE} Trans. Antennas Propag.},
  vol.~65, no.~4, pp. 1706--1713, Apr. 2017.

\bibitem{APM22AJJ}
J.~J. Adams, S.~Genovesi, B.~Yang, and E.~Antonino-Daviu, ``Antenna element
  design using characteristic mode analysis: {Insights} and research
  directions,'' \emph{{IEEE} Antennas Propag. Mag.}, vol.~64, no.~2, pp.
  32--40, Apr. 2022.

\bibitem{TAP20WJF}
J.~Wang \emph{et~al.}, ``Metantenna: {When} metasurface meets antenna again,''
  \emph{{IEEE} Trans. Antennas Propag.}, vol.~68, no.~3, pp. 1332--1347, Mar.
  2020.

\bibitem{APM22MD}
D.~Manteuffel, F.~H. Lin, T.~Li, N.~Peitzmeier, and Z.~N. Chen,
  ``Characteristic mode-inspired advanced multiple antennas: {Intuitive}
  insight into element-, interelement-, and array levels of compact large
  arrays and metantennas,'' \emph{{IEEE} Antennas Propag. Mag.}, vol.~64,
  no.~2, pp. 49--57, Apr. 2022.

\bibitem{TWC17PS}
S.~Park, A.~Alkhateeb, and R.~W. Heath, ``Dynamic subarrays for hybrid
  precoding in wideband mmwave {MIMO} systems,'' \emph{{IEEE} Trans. Wireless
  Commun.}, vol.~16, no.~5, pp. 2907--2920, May 2017.

\bibitem{TWC24JX}
X.~Jin \emph{et~al.}, ``A reconfigurable subarray architecture and hybrid
  beamforming for millimeter-wave dual-function-radar-communication systems,''
  \emph{{IEEE} Trans. Wireless Commun.}, vol.~23, no.~10, pp. 12\,594--12\,607,
  Oct. 2024.

\bibitem{TVT24WBW}
B.~Wang, H.~Li, and Z.~Cheng, ``Dynamic hybrid beamforming design for
  dual-function radar-communication systems,'' \emph{{IEEE} Trans. Veh.
  Technol.}, vol.~73, no.~2, pp. 2842--2847, Feb. 2024.

\bibitem{Tcom17SJH}
J.~Song, J.~Choi, and D.~J. Love, ``Common codebook millimeter wave beam
  design: {Designing} beams for both sounding and communication with uniform
  planar arrays,'' \emph{{IEEE} Trans. Commun.}, vol.~65, no.~4, pp.
  1859--1872, Apr. 2017.

\bibitem{TVT23LJW}
J.~Liu, X.~Li, T.~Fan, S.~Lv, and M.~Shi, ``Collaborative management of
  resource allocation and precoding for dual-mode networks,'' \emph{{IEEE}
  Trans. Veh. Technol.}, vol.~72, no.~8, pp. 10\,879--10\,893, Aug. 2023.

\bibitem{TWC17SO}
O.~Semiari, W.~Saad, and M.~Bennis, ``Joint millimeter wave and microwave
  resources allocation in cellular networks with dual-mode base stations,''
  \emph{{IEEE} Trans. Wireless Commun.}, vol.~16, no.~7, pp. 4802--4816, July
  2017.

\bibitem{JSTSP16YXH}
X.~Yu, J.-C. Shen, J.~Zhang, and K.~B. Letaief, ``Alternating minimization
  algorithms for hybrid precoding in millimeter wave {MIMO} systems,''
  \emph{IEEE J. Sel. Topics Signal Process.}, vol.~10, no.~3, pp. 485--500,
  Apr. 2016.

\bibitem{SPM14BE}
E.~Bj\"{o}rnson, M.~Bengtsson, and B.~Ottersten, ``Optimal multiuser transmit
  beamforming: {A} difficult problem with a simple solution structure,''
  \emph{{IEEE} Signal Process. Mag.}, vol.~31, no.~4, pp. 142--148, July 2014.

\bibitem{TSP11SQJ}
Q.~Shi, M.~Razaviyayn, Z.-Q. Luo, and C.~He, ``An iteratively weighted {MMSE}
  approach to distributed sum-utility maximization for a {MIMO} interfering
  broadcast channel,'' \emph{{IEEE} Trans. Signal Process.}, vol.~59, no.~9,
  pp. 4331--4340, Sep. 2011.

\end{thebibliography}

\end{document}